\documentclass[preprint,prb]{revtex4}
\usepackage{natbib}
\usepackage{amsmath,amssymb}
\usepackage{bm}
\usepackage{graphicx,color}
\newcommand{\rb }{{\bf r } }
\newcommand{\mat}[2]{|#1\rangle\langle#2|}
\newcommand{\dn}[1]{|#1\rangle}
\newcommand{\av}[1]{\langle#1\rangle}

\newcommand{\p}{\partial}

\begin{document}
\title{Proposals of nuclear spin quantum memory in group IV elemental and II-VI semiconductors}

\author{\"Ozg\"ur \c{C}ak{\i}r and Toshihide Takagahara}
\affiliation{Department of Electronics and Information Science,
Kyoto Institute of Technology,
Matsugasaki, Kyoto 606-8585 JAPAN}

\affiliation{ \\}

\affiliation{CREST, Japan Science and Technology Agency, 4-1-8 Honcho,
Kawaguchi,
Saitama 332-0012, JAPAN}

\date{\today}
\begin{abstract}
New schemes for the nuclear spin quantum memory are proposed based on a system composed of two electrons or one electron coupled to a single nuclear spin in isotopically purified group IV elemental and II-VI compound semiconductors. The qubit consists of the singlet state and one of the triplet states of two electrons or simply of an electron spin. These systems are free from the decoherence due to the nuclear dipole-dipole interaction and are advantageous for the long memory time. In the case of two electrons, the protocol for the quantum state transfer between the electron spin qubit and the nuclear spin qubit is based on the magnetic or electric field tuning of the singlet-triplet state crossing and on the hyperfine coupling supplemented with a well-defined scheme to initialize the nuclear spin. In the case of a single electron qubit, the quantum state transfer is driven by the hyperfine interaction itself without the need of the nuclear spin initialization. Many practical systems are considered, e.g., two electrons loaded on a Si or ZnSe quantum dot, a single electron charged state in a Si quantum dot doped with a P atom, a single electron charged $\;^{28}$Si quantum dot doped with an isotope atom of $\;^{29}$Si, and a localized electron system of Si:P and ZnSe:F in the bulk crystal. General aspects of these systems are investigated and a comparison of merits and demerits is made between the two-electron qubit and the single-electron qubit.

\end{abstract}
\maketitle

\section{Introduction}

Recently, a quantum media converter from a photon qubit to an
electron spin qubit was proposed for quantum
repeaters\cite{Yablonovitch03}. Quantum information can take several
different forms and it is preferred to be able to convert the information among
different forms. One form is the photon polarization
and another is the electron spin polarization. Photons are the most
convenient medium for sharing the quantum information between distant
locations. However, it is necessary to realize a quantum repeater in
order to send the information securely over a very long distance
overcoming the photon loss. A quantum repeater requires two
essential ingredients, namely, the quantum state transfer between a
photon and an electron spin \cite{Kosaka08,Kosaka09} and the
correlation(Bell) measurement between two electrons created by the
quantum state transfer from two different
photons\cite{Takagahara08}. Furthermore, it is desirable to have a
quantum memory to store the quantum state of the electron spin. The most suitable
medium for that purpose is the nuclear spins because of their
extremely long coherence time. 
It has been already proposed to use the collective nuclear spin state of the host medium, e.g.,
GaAs, as a quantum memory, employing the hyperfine (hf) interaction to transfer the electron spin state to the nuclear spins \cite{Taylor03a, Taylor03b}. However, in that proposal one has to achieve a high degree of nuclear polarization for the quantum memory of high fidelity, which has been prohibitive so far due to the low nuclear spin polarization achievable experimentally \cite{Reilly08,Tarucha07}.
In III-V compound semiconductors, all
the nuclei have a nonzero spin and the nuclear spin quantum memory
is necessarily subject to the decoherence induced by the nuclear
dipole-dipole interactions. Thus it is advantageous to employ a
system with a few nuclear spins like the nitrogen-vacancy (NV) center in diamond for
the quantum memory or register.  
The nitrogen spins of NV centers in diamond \cite{Dutt07} and phosphor spins in Si:P \cite{Lyon08} have been demonstrated to be promising as quantum memories. In these systems the degeneracies associated with nuclear spins are lifted by the hf interaction, enabling a selective addressing of nuclear spin states by external microwave or radio frequency (rf) fields. 

In view of these progress, we consider isotopically purified materials made of the
group IV elemental or II-VI compound semiconductors, such that the number of atoms with a nonzero nuclear spin can be reduced down to only one
and propose a few new schemes for the nuclear spin quantum memory in which the hf interaction 
itself drives the quantum state transfer (QST) between the electron spin qubit and the nuclear spin qubit. We study a system composed of two electrons or one electron and only one nuclear spin. Actual examples for the two-electron qubit are 
two electrons loaded in a $\;^{28}$Si quantum dot (QD) doped with an isotope atom of $\;^{29}$Si or in a ZnSe QD doped with an isotope atom of $\;^{77}$Se and a single-electron charged state 
in a Si QD doped with a P atom or in a ZnSe QD doped with a fluorine (F) atom. In these examples, the qubit is composed of the singlet state and one of the triplet states of two electrons and the QST is operated at the singlet-triplet crossing point. This crossing point can be approached by tuning the magnetic field or the electric field.
 The QST between the electron spin qubit and the nuclear spin qubit can be carried out by the hf interaction itself, reinforced with a well-defined scheme for the nuclear spin initialization which is based on the electron spin state measurement \cite{Cakir08}. A key requirement in this QST is that the singlet-triplet anticrossing gap should be much smaller than the hf interaction energy. We reveal that this requirement is satisfied favorably in the donor-bound electron system but not in the delocalized electron system in a QD.
 
On the other hand, for the single-electron qubit, practical examples are a single-electron charged $\;^{28}$Si QD doped with an isotope atom of $\;^{29}$Si, a single-electron charged ZnSe QD doped with an isotope atom of $\;^{77}$Se, and a localized electron system of Si:P and ZnSe:F in the bulk crystal. Here the electron spin plays the role as a qubit and the single nuclear spin of an isotope atom or a donor atom plays the role as a quantum memory. It is advantageous that the relative magnitude between the spin-orbit interactions and the hf coupling energy is not relevant in this QST. A more important feature in this QST is that the nuclear spin initialization is not necessary.

The paper is organized as follows.
We first present the results on the magnetic field tuning of the singlet-triplet state crossing of two electrons loaded on a single QD and of two electrons in a single-electron charged QD doped with a donor atom. We discuss the effects of the spin-orbit interactions which induce the singlet-triplet state anticrossing and become obstacles to our QST protocols. Then we discuss the hyperfine interaction and propose a few protocols for the QST between the electron spin qubit and the nuclear spin qubit. We also investigate the system composed of a single electron and a single nuclear spin and present a QST protocol between them, because the single-electron qubit is more fundamental as a building block of devices for the quantum information processing.
Finally, we make a comparison of merits and demerits between the two-electron qubit and the single-electron qubit and conclude that the donor-bound single electron system is preferable with respect to the fast QST operation, no need of the nuclear spin initialization and the irrelevance to the relative magnitude between the spin-orbit interaction and the hyperfine interaction.

\section{Magnetic field tuning of the singlet-triplet state crossing of two electrons in a quantum dot}

\subsection{Energy spectrum of a pair of electrons in a Si or ZnSe quantum dot}

First we consider a pair of electrons in a Si QD. In bulk Si, the conduction band minima have a six-fold
degeneracy, which makes it difficult to define a robust electronic
qubit. However, this degeneracy is lifted in Si quantum well (QW)
structures. The strain in the lateral direction ($x$ and $y$) lifts up the four conduction band minima along the $x$ and $y$ directions by about 100 meV \cite{Horiguchi}. Additionally, the confinement in the growth direction couples the lowest two conduction band
minima along the $z$ direction, removing the degeneracy
completely \cite{Friesen07}. At the conduction band minima along the
$z$ direction, the Bloch functions, which are separated by an energy difference about $1.5$meV \cite{Goswami07}, are given by
\begin{eqnarray}
 \psi_+(\rb)=\sqrt{2}\cos(k_0z+\phi)|u_{k_0}(\rb)| \;, \\
\psi_-(\rb)=\sqrt{2}\sin(k_0z+\phi)|u_{k_0}(\rb)| \;,
 \end{eqnarray}
where $(0, 0, \pm k_0) \; (k_0\simeq 0.85 \pi/a)$ is the wave
vector at the band minima along the $z$ direction with the lattice constant $a$
of Si, $\phi$ a phase factor related to the valley mixing, and $u_{k_0}(\rb)$ is the periodic part of the Bloch function at the band minima normalized in the unit cell volume. The lower-energy one of $\psi_+$ and $\psi_-$ is determined depending on the QW thickness \cite{Friesen07}. The actual wavefunction is given by the product of the lowest energy Bloch function and the envelope function $F(\rb)$ satisfying the
effective mass equation.   Under a magnetic field along the $z$ direction the effective mass equation for the envelope function is given by
\begin{eqnarray}
&&\Bigl[\frac{1}{2m_t}\bigl((p_x+\frac{e}{c} A_x)^2+ (p_y+\frac{e}{c} A_y)^2\bigr)+\frac{p_z^2}{2m_l}+U(x,y)+V(z)\Bigr ] F(\rb)=
\epsilon F(\rb)  \label{eme} \\
&&{\rm with} \;\; U(x,y)=\frac{1}{2}m_t \omega_0^2(x^2+y^2) \;, \; {\bf A}=\frac{B}{2}(-y,x,0) \;, \label{env}
\end{eqnarray}
where $U$ represents the harmonic confinement in the lateral direction, $V$ is an additional potential in the $z$ direction describing, e.g., the electrostatic confinement, $m_t(=0.19\;m_0, m_0$ being the free electron mass) and $m_l(=0.92\;m_0)$ are the transverse and
longitudinal effective masses, respectively. The Zeeman energy term does not appear in Eq. (\ref{eme}) because $F(\rb)$ represents only the orbital part. But the Zeeman energy is included in the calculation of the energy level structure.

\begin{figure}
 \includegraphics[angle=0]{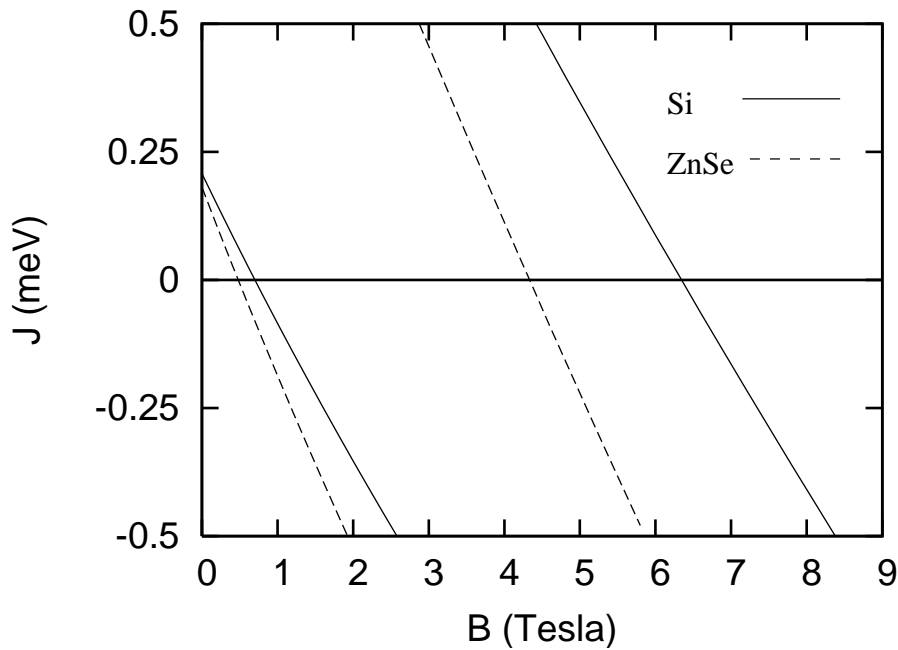}
\caption{The exchange energy $J=E({\rm triplet})-E({\rm singlet})$, namely, the energy difference between the ground singlet state and the excited triplet state in a circularly symmetric quantum dot is plotted as a function of the magnetic field strength. The intercept with the line $J=0$ represents the singlet-triplet crossing point. The solid (dashed) line corresponds to the Si (ZnSe) quantum dot. 
Curves on the left (right) are plotted for $\hbar \omega_0=1 (5)$ meV. \label{fig_spect}\label{fig_znse}}
\end{figure}

For a vanishing magnetic field, the ground state of two electrons is the spin singlet state\cite{Mattis}.
However, in the presence of a magnetic field, the energy spectrum does not necessarily follow the Lieb-Mattis theorem.  The magnetic confinement and the Coulomb interaction together lead
to crossings in the energy spectrum. In Fig.\ref{fig_spect}, the exchange energy $J=E({\rm triplet})-E({\rm singlet})$, namely, the energy difference between the ground singlet state and the excited triplet state of two electrons, is plotted
for two values of the lateral confinement energy $\hbar\omega_0$. The point at $J=0$ indicates the singlet-triplet crossing point.
When electrons are confined weakly in the lateral direction, the effect of the magnetic field on the orbital motion begins to appear at the weak field, influencing the exchange energy between two electrons. Conversely, under the strong confinement in the lateral direction, the effect of the magnetic field on the exchange energy between two electrons becomes manifest at the strong field, shifting the singlet-triplet crossing point to a higher magnetic field. These qualitative features can be confirmed in Fig. 1. 

\begin{figure}
 \includegraphics[angle=0,width=13cm]{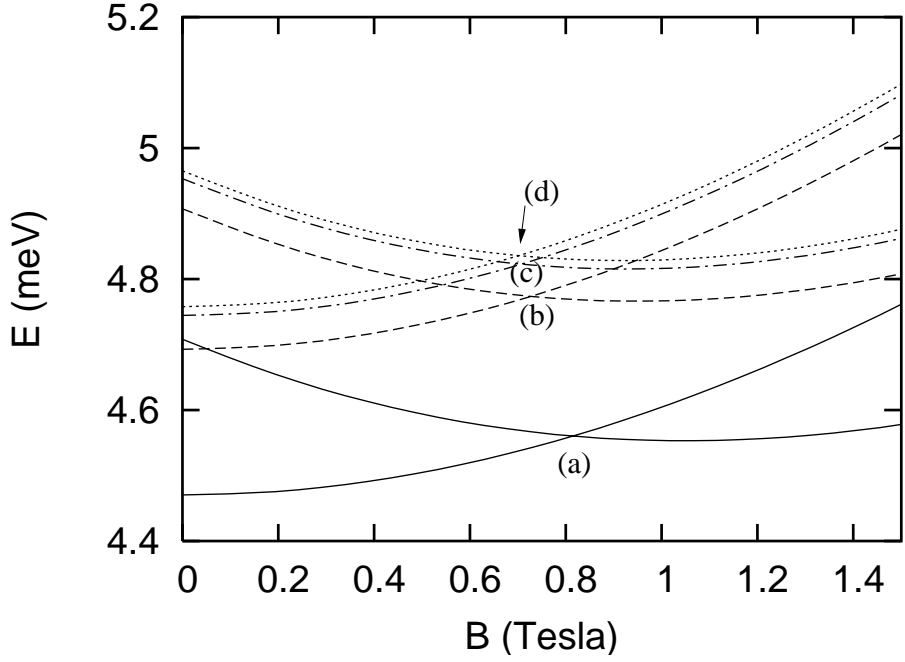}
\caption{Energy level crossing between the ground singlet (growing curve with the magnetic field) and the triplet (decreasing curve with the magnetic field) states in a Si QD with the fixed lateral confinement energy $\hbar \omega_0=1$ meV and the additional confinement energy in the growth direction, $\hbar \omega_z=$ (a) 1 (solid curves), (b) 5 (dashed curves), (c) 25 meV (dot-dashed curves) and (d) $\infty$ (dotted curves). As $\hbar \omega_z$ is increased, the level crossing occurs at a higher energy and at a smaller magnetic field. In this plotting the confinement energy in the $z$ direction is subtracted for the ease of comparison.  \label{finw} }
\end{figure}

Now we discuss the effect of the additional potential in the $z$ direction produced by, e.g., electrical gates. To facilitate the arguments, that effect is taken into account by assuming a harmonic potential for $V(z)$ in Eq. (\ref{eme}) as
\begin{equation}
V(z)=\frac{1}{2} m_l \omega^2_z z^2 \;,
\end{equation}
where $\hbar \omega_z$ is the harmonic confinement energy. In general, the spatial extension in the growth direction enhances the three-dimensional character of the electron motion and
leads to the reduction in the Coulomb energy and to the weaker dependence of the orbital motion on the magnetic field, pushing the singlet-triplet crossing point to higher magnetic
fields. In Fig.\ref{finw}, the effect of the additional potential on the energy spectrum is exhibited.
As the confinement energy $\hbar \omega_z$ in the $z$ direction is increased, the singlet-triplet crossing 
point moves to higher energies because of the increase in the Coulomb energy and shifts to weaker magnetic fields because the two-dimensional character of the electron motion is enhanced and the orbital motion becomes more susceptible to the magnetic confinement. 

 As an example of the direct-gap material, we consider a ZnSe QD. Contrary to Si, the bulk ZnSe has the conduction band minimum at the $\Gamma$ point with an isotropic effective mass. The conduction band electron can be described by the wavefunction:
\begin{eqnarray}
 \Psi(\rb)= u_{0}({\bf r}) \;F({\bf r}),
\end{eqnarray}
where $u_{0}({\bf r})$ is the Bloch function at the $\Gamma$ point and the envelope function $F({\bf r})$ satisfies Eq. (\ref{eme}) with a modification of $m_t=m_l$.
Thus the physics of
the singlet-triplet crossing is the same as discussed above. The crossing behaviors are shown in Fig.\ref{fig_znse}, employing the
 effective electron mass $m_t=m_l=0.16\; m_0$ and the dielectric constant $\kappa=9.1$ \cite{Bornstein17ab}.
 

\begin{figure}
\includegraphics[width=12cm,clip]{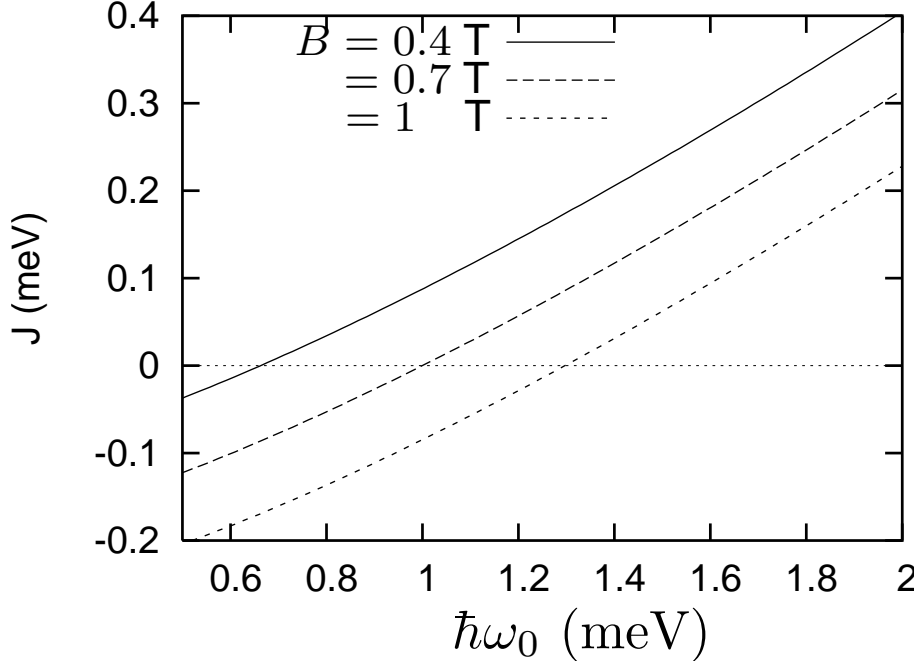}
\caption{The exchange energy $J$, i.e., the singlet-triplet energy difference, is plotted as a function of
the lateral confinement energy $\hbar \omega_0$ for the magnetic field strength $B=$ 0.4, 0.7 and 1 Tesla from top to bottom.} \label{fig_conf}
\end{figure}

The singlet-triplet crossing of a pair of electrons on a GaAs QD was observed by the gate voltage
tuning\cite{Kyriakidis02}. The gate voltage controls the number of electrons in the QD
as well as the shape of the QD potential electrostatically, enabling the ground state tuning.
In Fig. \ref{fig_conf}, the
dependence of the exchange energy $J$ on the lateral confinement energy $\hbar \omega_0$ is depicted for a few values of the magnetic field strength.
 When the lateral confinement is increased, the orbital energies increase directly proportional to the confinement energy. At the same time, the spatial overlap between electron orbitals is enhanced and consequently the direct and exchange Coulomb energies increase. This leads to the increase of $J$ in proportion to the lateral confinement energy $\hbar \omega_0$, as seen in Fig. \ref{fig_conf}.
The lateral confinement energy or the potential curvature can be controlled by tuning the voltages on several gates. Thus, either by electrically modifying the lateral confinement potential or by changing the magnetic field, one can tune the singlet-triplet crossing. 

\subsection{Effects of spin-orbit interactions}

So far we have neglected the effect of the spin-orbit(SO) coupling. But this effect should be examined because the SO coupling mixes the spin states through the orbital state mixing and affects the singlet-triplet crossing \cite{Golovach}. If the spin-orbit interaction has a non-zero matrix element between the singlet (S) and triplet (T) states, the level crossing turns into the level anticrossing which becomes an obstacle to the QST driven by the hf interaction. In order to achieve securely the electron-nuclear QST, it is required that the matrix element of the spin-orbit interaction is much smaller than that of the hf interaction. In the following we will examine this requirement.

The spin-orbit coupling for the conduction band electron in the linear approximation with respect to the momentum operator is given as
\begin{equation}
 V_{SO}=a_R(\sigma_x p_y-\sigma_y p_x)+ a_D(\sigma_x p_x-\sigma_y p_y) \;, \label{so}
\end{equation}
where the first term is the Rashba SO coupling due to the structural inversion asymmetry\cite{Rashba84a} and the second term is the Dresselhaus term arising from the bulk inversion asymmetry\cite{Dresselhaus55}.  Here the momentum ${\bf p}$ is meant by the gauge invariant kinetic momentum $-i\hbar{\bf \nabla}+e{\bf A}/c$ with the vector potential ${\bf A}$, if the system is under a magnetic field. In order to discuss the S-T anticrossing, it is convenient to eliminate the SO coupling by applying a unitary transformation \cite{Aleiner} to the Hamiltonian as discussed in Appendix A. The original single electron Hamiltonian is given by
\begin{equation}
H=\frac{1}{2m} {\bf p}^2+ a_R (\sigma_x p_y-\sigma_y p_x)+ a_D (\sigma_x p_x-\sigma_y p_y) +\frac{1}{2}g_e \mu_B \bm{B}\cdot \bm{\sigma} \;,
\end{equation}
where $g_e$ is the electron $g$-factor.
By rotating the coordinate system by an angle $\pi/4$ in the $xy$ plane, namely, by introducing a new coordinate system ($\xi, \eta, z$) defined by ${\bf e}_{\xi}=(1,1,0)/\sqrt{2}$, ${\bf e}_{\eta}=(-1,1,0)/\sqrt{2}$, ${\bf e}_z=(0,0,1)$, the Hamiltonian is rewritten as
\begin{eqnarray}
&&H=\frac{1}{2m} {\bf p}^2 -p_{\xi}\sigma_{\eta}(a_D + a_R)+p_{\eta}\sigma_{\xi}(a_R - a_D) +\frac{1}{2} g_e \mu_B (B_{\xi}\sigma_{\xi}+ B_{\eta}\sigma_{\eta} +B_z \sigma_z) \nonumber \\
&&=\frac{1}{2m} \left(p_{\xi}- \frac{\hbar}{\lambda_{\xi}}\sigma_{\eta} \right)^2 + \frac{1}{2m} \left(p_{\eta}+ \frac{\hbar}{\lambda_{\eta}}\sigma_{\xi} \right)^2 + \frac{1}{2m} p^2_z \nonumber \\
&& +\frac{1}{2} g_e \mu_B (B_{\xi}\sigma_{\xi}+ B_{\eta}\sigma_{\eta} +B_z \sigma_z)
-m(a^2_D +a^2_R)  \label{aeqq19} \\
{\rm with} \;\; && \lambda_{\xi}=\frac{\hbar}{m(a_R+a_D)} \;,\; \lambda_{\eta}=\frac{\hbar}{m(a_R-a_D)} \;,
\end{eqnarray}
where the last constant term in Eq. (\ref{aeqq19}) will be omitted. Now, in order to eliminate the original spin-orbit 
coupling, we apply the unitary transformation:
\begin{equation}
\tilde{H}=U^{\dag} \;H\;U \quad {\rm with} \; U=\exp[i\frac{\xi}{\lambda_{\xi}}\sigma_{\eta} - i\frac{\eta}{\lambda_{\eta}}\sigma_{\xi}] \;.
\end{equation}
Then we have
\begin{eqnarray}
&&\tilde{H}=\frac{1}{2m} {\bf p}^2 +\frac{1}{2} g_e \mu_B (B_{\xi}\sigma_{\xi}+ B_{\eta}\sigma_{\eta} +B_z \sigma_z) \nonumber \\
&&+ g_e \mu_B \left[ \left(\frac{\xi}{\lambda_{\xi}} \sigma_{\xi}+\frac{\eta}{\lambda_{\eta}} \sigma_{\eta}\right) B_z - \left(\frac{\xi}{\lambda_{\xi}} B_{\xi} +\frac{\eta}{\lambda_{\eta}} B_{\eta} \right) \sigma_z \right] \nonumber \\
&&+\frac{\hbar}{m \lambda_{\xi} \lambda_{\eta}} \left[ -L_z \sigma_z +\frac{\sigma_{\xi}}{\lambda_{\xi}}(-2\xi^2 p_{\eta}+\eta \{\xi, p_{\xi}\}) + \frac{\sigma_{\eta}}{\lambda_{\eta}}(2\eta^2 p_{\xi}- \xi \{\eta, p_{\eta}\}) \right] \;,
\end{eqnarray}
where $\{A, B\} \equiv AB + BA$, the first line represents the single electron Hamiltonian under a magnetic field without the SO coupling, the second line the SO induced Zeeman interaction $H^{SO}_Z$ and the third line 
stands for the renormalized SO interaction $H_{ren}$. The energy level structure of $\tilde{H}$ is exactly the same as that of the original Hamiltonian $H$. Thus we can discuss the S-T crossing behavior based on the transformed Hamiltonian $\tilde{H}$.
After this unitary transformation, the original SO interaction is eliminated but new terms appear with a smallness parameter defined by $\varepsilon \equiv \ell_t/\lambda_{SO}$ which is typically about $10^{-3}$, where $\ell_t$ is the lateral extent of the electron wavefunction and $\lambda_{SO}$ represents symbolically the SO length $\lambda_{\xi}$ and $\lambda_{\eta}$. Among the newly appeared terms, the SO induced Zeeman interaction $H^{SO}_Z$ is the most dominant term which is of the first order in the smallness parameter $\varepsilon$. However, the matrix element of this term between the singlet state $|S\rangle$ and the triplet states $|T_{\pm}\rangle$, i.e., $\langle S|H^{SO}_Z|T_{\pm}\rangle$ can be made to vanish by tuning the direction of the magnetic field as shown in Appendix A. When one of the SO coupling constants $a_D$ and $a_R$ is much larger than the other, the perpendicular (z-directed) magnetic field is most favorable. However, the actual magnitude of the S-T anticrossing gap is determined by the higher order perturbation terms of $H^{SO}_Z$ and contributions from the renormalized SO term $H_{ren}$. The typical magnitude is about $10^{-12} \sim 10^{-11}$ eV as shown in Appendix A. 

So far we discussed the linear-in-momentum SO coupling. 
However, the cubic-in-momentum SO term is present in general and more detailed arguments are necessary. As discussed in Appendix A, the Dresselhaus SO term can be decomposed as
\begin{eqnarray}
&&V_D=V_D^{(1)}+V_D^{(3)} \\
{\rm with} \; && V_D^{(1)}= \langle \phi(z)|V_D|\phi(z)\rangle= \gamma \langle \phi(z)| k^2_z |\phi(z)\rangle (- k_x \sigma_x + k_y \sigma_y) \;, \\
&&V_D^{(3)}= \gamma (k_x k^2_y \sigma_x -k_y k^2_x \sigma_y) \;,
\end{eqnarray}
where $\phi(z)$ is the ground state orbital in the $z$-direction,
the linear term $V_D^{(1)}$ is already included in the above $V_{SO}$ and $V_D^{(3)}$ denotes the cubic-in-momentum SO term. Unfortunately, this $V_D^{(3)}$ cannot be eliminated by the unitary transformation as shown below
\begin{eqnarray}
&&U^{\dag}\;V_D^{(3)} \;U = V_D^{(3)} + \frac{\gamma}{2\hbar^2 \lambda_{\xi}} (p^2_{\eta} -3 p^2_{\xi})
-\frac{\gamma}{2\hbar^2 \lambda_{\eta}} (p^2_{\xi} -3 p^2_{\eta}) \nonumber \\
&&+ \frac{\gamma}{4\hbar^3} \Bigl[ \frac{2i}{\lambda_{\xi}}(-\hbar\{p_{\xi}, p_{\eta}\} +i(\{p^2_{\xi}, p_{\eta}\} -2 p^3_{\eta}) \xi )
 + \frac{2i}{\lambda_{\eta}}(-\hbar\{p_{\xi}, p_{\eta}\} +i(\{p^2_{\eta}, p_{\xi}\} -2 p^3_{\xi}) \eta ) \Bigr] \sigma_z + \cdots \;, \nonumber \\
&&= V_D^{(3)} + V_{D}^{(3)ren} + \cdots \;,
\end{eqnarray}
where the newly appeared terms denoted by $V_{D}^{(3)ren}$ are smaller in 
magnitude than the original terms due to the smallness parameter $\varepsilon \equiv \ell_t/\lambda_{SO} \sim 10^{-3}$. Furthermore, $V_{D}^{(3)ren}$ does not contribute to the matrix element between the singlet state $|S\rangle$ and the triplet states $|T_{\pm}\rangle$, because this does not change the magnetic quantum number.
Thus we have to discuss the effect of $V_D^{(3)}$ on the S-T anticrossing gap .
As shown in Appendix A, the matrix element $\langle S|V_D^{(3)}|T_{\pm} \rangle$ vanishes, if the orbital excited state associated with the $T_{\pm}$ is chosen as
\begin{equation}
e^{\mp}_{rel}(x,y)=\frac{1}{\ell_t} (x \mp iy) \; g_{rel}(x,y)\quad {\rm with} \quad \;g_{rel}(x,y)=\frac{1}{\ell_t \sqrt{\pi}} \exp[-\frac{1}{2\ell^2_t}(x^2+y^2)] \;.
\end{equation}
The higher order perturbation terms concerning $V_D^{(3)}$ and the contributions from $V_{D}^{(3)ren}$ are estimated in Appendix A. The typical magnitude of the S-T anticrossing gap is about $10^{-12} \sim 10^{-11}$ eV, which is of the same order of magnitude as that for the linear-in-momentum SO coupling. Consequently, we can summarize that the S-T anticrossing gap is comparable to the hyperfine coupling energy as discussed in Sec. IV for the case of two delocalized electrons in a QD with an isotope atom of the host material. This situation is unfavorable for our protocol of the nuclear spin quantum memory which requires that the S-T anticrossing gap is much smaller than the hyperfine coupling energy.

However, we consider also the case of the single electron charged QD with a donor impurity having the nuclear spin in the next Section and find that the S-T anticrossing gap can be much smaller than the hyperfine coupling energy. Thus our protocol of the nuclear spin quantum memory will be fully effective in this case.

\section{Energy spectrum of a single electron charged QD doped with a donor atom}

\begin{figure}
\includegraphics[width=15cm,clip]{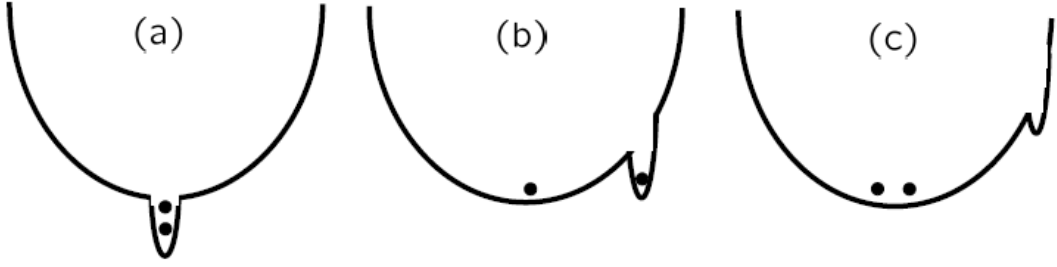} 
\caption{Schematic configuration of two electrons in a single electron charged QD doped with a donor atom. The large (small) parabola represents the QD confinement potential (Coulomb potential by an ionized donor). A black dot indicates an electron. Three regimes are depicted: (a) two electrons are bound to the donor, (b) one electron is bound to the donor and the other is delocalized over the QD, and (c) both electrons are delocalized over the QD. } \label{donor}
\end{figure}

Here we investigate the energy spectrum of a singly charged QD doped with a donor. The extra electron will hybridize with the donor electron so as to form the singlet-triplet states.
We can envision three regimes as shown in Fig. \ref{donor}, namely, (a) two electrons are bound to the donor, (b) one electron is bound to the donor and the other is delocalized over the QD, and (c) both electrons are delocalized over the QD.
These regimes depend on the position of the donor atom. When the donor is doped at the central part of the QD, two electrons are bound strongly by the donor atom, whereas when the donor is doped at the peripheral region of the QD, two electrons are trapped weakly by the QD confinement potential. In the intermediate case, the donor atom and the QD potential provide the same order of confinement for a trapped electron and there occurs the situation that one electron is bound to the donor and the other is delocalized over the QD. 
For cases (a) and (c), it is already known that there occurs a singlet-triplet level crossing induced by the magnetic field and the Coulomb interaction.\cite{Cakir08,Henry74,Hujaj00}
In this section, we focus on the case (b) and confirm that the singlet-triplet level crossing occurs at appropriate values of the distance of the donor atom from the center of the QD. 

The Hamiltonian for two electrons in a negatively charged donor-doped QD is given by
\begin{equation}
 H=h(\rb_1)+h(\rb_2)+\frac{e^2}{\kappa|\rb_1-\rb_2|} \;. \label{hamilt}
\end{equation}
Here $h$ is the single-particle Hamiltonian given by
\begin{eqnarray}
&&h=T+h_Z+ V_a+V_b \\
{\rm with} && T=\frac{1}{2m}({\bf p}+\frac{e}{c}{\bf A})^2 \;, \; h_Z=\frac{1}{2} g_e \mu_B B \sigma_z \;, \; V_a=\frac{1}{2}m(\omega_0^2 \rho^2+\omega_z^2 z^2) \;, \nonumber \\
&& V_b=-\frac{e^2}{\kappa |\rb-d\hat{x}|} \;, \; \rho^2=x^2 + y^2 \;, \; {\bf A}=\frac{B}{2}(-y,x,0) \;,
\end{eqnarray}
where $T$ is the kinetic energy part including the effect of a magnetic field applied along 
the $z$-direction, $h_Z$ the Zeeman energy with the electron $g$-factor $g_e$ and the Bohr magneton $\mu_B$, $V_a$ the confinement potential by the QD, 
$V_b$ the Coulomb potential by an ionized donor displaced by a distance $d$ from the center of the QD and $\hat{x}$ is the unit vector in the $x$-direction.
 
In the Heitler-London approach, the orbital wavefunctions for the spin-singlet ($\Psi_+$) state and the spin-triplet ($\Psi_-$) state are approximated as
\begin{eqnarray}
&&\Psi_\pm=\frac{1}{\sqrt{2(1\pm |S|^2)}}\left(\phi_a(\rb_1)\phi_b(\rb_2) \pm 
\phi_a(\rb_2) \phi_b(\rb_1) \right) \\
{\rm with} && S=\av{\phi_a|\phi_b} \;,
\end{eqnarray}
where $\phi_a$ and $\phi_b$ are appropriate single-particle wavefunctions specified later. 
Then we have 
\begin{align}
 E_\pm&=\langle\Psi_\pm|H|\Psi_\pm\rangle\\
&=\frac{1}{1\pm|S|^2}\bigl(e_a+e_b +\av{\phi_a|V_b|\phi_a}+\av{\phi_b|V_a|\phi_b}+V_d\bigr) \nonumber\\ 
&\pm\frac{1}{1\pm|S|^2}\bigl(\av{\phi_a|h|\phi_b}S^*+\av{\phi_b|h|\phi_a}S+V_X\bigr) \label{epm}\\
{\rm with} \;\; &V_d=\av{\phi_a(\rb_1)\phi_b(\rb_2)|\frac{e^2}{\kappa|\rb_1-\rb_2|}|\phi_a(\rb_1)\phi_b(\rb_2)} \;, \\
&V_X=\av{\phi_a(\rb_1)\phi_b(\rb_2)|\frac{e^2}{\kappa|\rb_1-\rb_2|}|\phi_a(\rb_2) \phi_b(\rb_1)} \;, \\
&e_a=\av{\phi_a|T+V_a|\phi_a} \;, \; e_b=\av{\phi_b|T+V_b|\phi_b} \;,
\end{align}
where $V_d (V_X)$ is the direct (exchange) Coulomb energy. 
When only the $\phi_a$ can be chosen to be an exact eigenstate:
\begin{equation} 
(T+V_a)\phi_a=e_a\phi_a \;,
\end{equation}
the exchange energy, namely, the energy difference between the singlet and the triplet states, is calculated as
\begin{eqnarray}
 J=E_--E_+ &&=\frac{2|S|^2}{1-|S|^4}\bigl((e_b-e_a)+ \av{\phi_a|V_b|\phi_a}+\av{\phi_b|V_a|\phi_b}+V_d\bigr) \nonumber \\
&& -\frac{2}{1-|S|^4}\bigl(\av{\phi_a|V_b|\phi_b}S^*+\av{\phi_b|V_b|\phi_a}S+V_X \bigr) +g_e \mu_B B m_z \;, \label{j}
\end{eqnarray}
where the Zeeman energy is included with the magnetic quantum number $m_z$ of the triplet state

Now, $\phi_a$ is chosen as the exact ground state in the QD confinement potential $V_a$:
\begin{eqnarray}
&& \phi_a(\rb)=\frac{1}{\pi^{3/4}} \frac{1}{\ell_t\sqrt{\ell_z}} \exp[-\frac{\rho^2}{2 \ell_t^2}-\frac{z^2}{2 \ell_z^2}]   \label{phia} \\
{\rm with} &&\ell_t=\sqrt{\frac{\hbar}{m\Omega}} \;, \;\ell_z=\sqrt{\frac{\hbar}{m\omega_z}} \;, \;
\Omega=\sqrt{\omega_0^2+\frac{\omega_c^2}{4}} \;,\; \omega_c=\frac{|e|B}{mc} \;, \label{length}
\end{eqnarray}
where $\omega_c$ is the cyclotron frequency and the eigenenergy $e_a$ is $\hbar \Omega+\hbar \omega_z/2$.
 On the other hand, the donor-bound eigenstates are calculated variationally in the Gaussian functions to facilitate the following calculations. The ground state is approximated as
\begin{equation}
\psi(r)=\sqrt{\frac{\alpha\beta^{1/2}}{\pi^{3/2}a_B^3}}\exp[-\frac{\alpha\rho^2}{2a_B^2}-\frac{\beta z^2}{2a_B^2}] \;, \label{psi}
\end{equation}
where $\alpha$ and $\beta$ are the variational parameters and the corresponding energy is calculated as
\begin{eqnarray}
E[\psi]&&=\av{\psi|T-\frac{e^2}{\kappa |\rb|}|\psi}\\
&&=\left\{\begin{array}{l}
\frac{R}{2}\Bigl\{ \alpha+\frac{\beta}{2}+\frac{\gamma^2}{\alpha}-\frac{2\sqrt{\beta}}{\sqrt{\pi}\sqrt{1-\beta/\alpha}}
\ln|\frac{1+\sqrt{1-\beta/\alpha}}{1-\sqrt{1-\beta/\alpha}}| \Bigr\} \quad {\rm for} \; \alpha \geq \beta \;, \\
\frac{R}{2}\Bigl\{ \frac{3}{2}\alpha+\frac{\gamma^2}{\alpha} -4\sqrt{\frac{\alpha}{\pi}}
\Bigr\} \quad {\rm for} \; \beta = \alpha \;, \\
\frac{R}{2}\Bigl\{ \alpha+\frac{\beta}{2}+\frac{\gamma^2}{\alpha}-\frac{4\sqrt{\beta}}{\sqrt{\pi} \sqrt{\beta/\alpha-1}}\arctan[\sqrt{\beta/\alpha-1}]
\Bigr\} \quad {\rm for} \; \beta \geq \alpha \;,
\end{array}\right. \label{energy}\\
{\rm with} && R=\frac{m e^4}{\hbar^2\kappa^2},\;a_B=\frac{\hbar^2\kappa}{me^2},\;\gamma=\frac{\hbar\omega_c}{2 R} \;.
\end{eqnarray}
In the absence of a magnetic field there is no preferential direction in the donor-bound state and thus $\alpha=\beta$. 
Under a magnetic field along the $z$-direction, the squeezing of the wavefunction occurs in the transverse direction and $\alpha\geq\beta$.

The eigenstate of an electron in the presence of a magnetic field and an ionized donor at a position shifted from the origin is determined by
\begin{equation}
\Bigl [T -\frac{e^2}{\kappa |\rb-d\hat{x}|} \Bigr]\phi_b(\rb)=e_b\phi_b(\rb) \;.\\ 
\end{equation}
The shift of the position of the donor atom can be taken into account by introducing a phase shift as
\begin{equation}
\phi_b(\rb) \cong \exp[-i\frac{eBd}{2\hbar c}y] \;\psi(\rb-d \hat{x}) \;, \label{phib}
\end{equation}
where $\psi$ is the variational wavefunction in Eq. (\ref{psi}) and the corresponding energy is given by Eq. (\ref{energy}). The relevant integrals appearing in the expression of $J$ in Eq. (\ref{j}) are given in Appendix B.

\begin{figure}
$\begin{array}{cc}
 \includegraphics[width=0.5\linewidth]{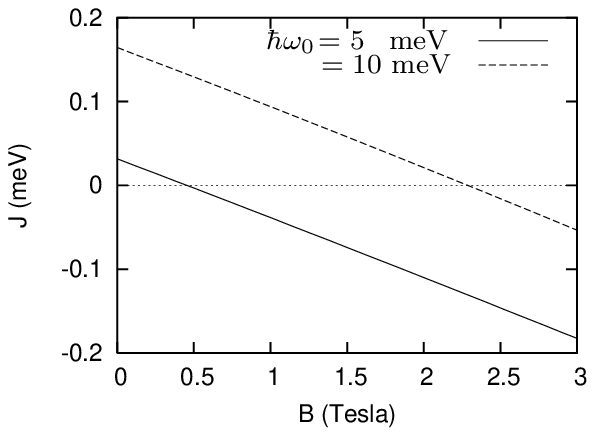} &
 \includegraphics[width=0.5\linewidth]{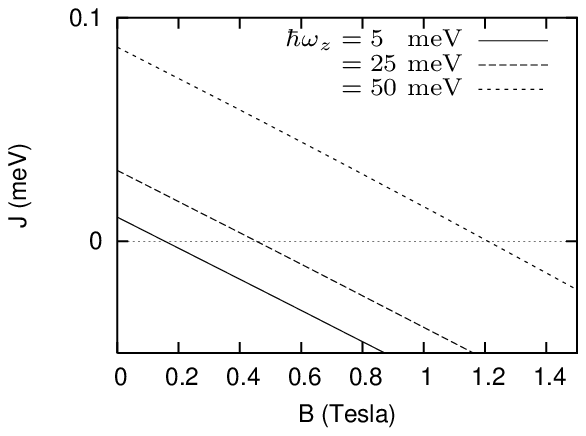}\\ (a) & (b)
\end{array}$
\caption{The energy difference between the triplet $(T_+)$ and the singlet $(S)$ states for a donor doped ZnSe QD is plotted as a function of the magnetic field. The left panel corresponds to the case (a) $\hbar\omega_0=5, 10$ meV and $\hbar\omega_z=25$ meV with a donor at the distance $d=3\; \ell_0 (\ell_0= \sqrt{\hbar/(m\omega_0)})$ from the center of the QD. The right panel exhibits the case (b)  $\hbar\omega_0=5$ meV and $\hbar\omega_z=5, 25, 50$ meV with a donor at the distance $d=3\; \ell_0$ from the center of the QD.} \label{stwz}
\end{figure}

 The Bohr radius $a_B$ of the donor electron is about a few nm, whereas the confinement length in Eq. (\ref{length}) is about 10-20 nm for $\hbar \omega_0=1 \sim 5$meV. Thus the spatial overlap between $\phi_a$ and $\phi_b$ is expected to be small and correspondingly the exchange integral $V_X$ is small, leading to the weak dependence of the singlet-triplet splitting energy $J$ on the magnetic field. In this range the Zeeman energy is comparable to or larger than $V_X$ and determines the magnetic field at which the singlet-triplet crossing occurs.
The regime (b), where one electron is bound to the donor and the other electron is delocalized over the QD, corresponds to the situation that the harmonic confinement potential is compensated by the donor binding energy, namely,
\begin{equation}
\frac{1}{2} m \omega^2_0 d^2 \sim \frac{1}{2} R \longrightarrow d \sim \frac{R}{\hbar \omega_0} a_B \;.
\end{equation}
For ZnSe, $a_B$=3.0 \;nm, $R$=52.6 meV and thus for $\hbar \omega_0$=5 (10) meV, the regime (b) corresponds to $d \sim$ 31.6 (15.8) nm. The harmonic confinement length $\ell_0=\sqrt{\hbar/(m \omega_0)}$ is estimated as 9.76 (6.90) nm and $d/\ell_0=$ 3.2 (2.3). The S-T$_+$ crossing behaviors
are exhibited in Fig. \ref{stwz} for ZnSe QDs with $g_e=1.2$. Figure \ref{stwz}(a) exhibits the case of $\hbar\omega_0=5, 10$ meV and $\hbar\omega_z=25$ meV with a donor at the distance $d=3 \;\ell_0$ from the center of the QD. A clear S-T$_+$ level crossing occurs at an easily available magnetic field strength. 
With increasing $\hbar\omega_z$, the wavefunction is squeezed along the z-direction and the spatial overlap between wavefunctions and accordingly the exchange Coulomb energy is decreased. In order to increase the exchange Coulomb energy, more tighter magnetic confinement of electrons is required, pushing the S-T$_+$ level crossing to the higher magnetic field for the stronger confinement in the $z$-direction. 
Figure \ref{stwz}(b) illustrates these behaviors for $\hbar\omega_0=5$ meV and $\hbar\omega_z$=5, 25 and 50 meV with a donor at the distance $d=3 \;\ell_0$. In this case also, the S-T$_+$ level crossing occurs at around 1 Tesla.

Finally, we discuss the effect of the spin-orbit (SO) coupling on the S-T$_{\pm}$ anticrossing gap in the regime (b) of the donor-doped QD. 
The detailed arguments are developed in Appendix A. A unitary transformation is applied in order to eliminate the linear-in-momentum SO coupling. Then the most dominant term is the SO induced Zeeman interaction $H^{SO}_Z$ as discussed in Sec. II B, which is of the first order in the smallness parameter defined by $\varepsilon=\ell_t/\lambda_{SO}$. Unfortunately, the matrix element $\langle S|H^{SO}_Z|T_{\pm} \rangle$ cannot be made to vanish by tuning the direction of the magnetic field. Thus the magic angle tuning is not possible here. Instead, in order to reduce the matrix element, we have to use the group IV elemental semiconductors in which the Dresselhaus SO terms are absent and the magnitude of the Rashba SO term might be reduced very much by careful tuning of the strain fields and the electric field. If we can achieve $\ell_t/\lambda_{\xi}, \;\ell_t/\lambda_{\eta} \sim 10^{-4}$, the above matrix element would be about $10^{-8}$ eV. At the same time, the contribution from the higher order perturbation series with respect to $H^{SO}_Z$ is much smaller. The contributions from the renormalized SO interaction $H_{ren}$, the cubic-in-momentum SO interaction $V^{(3)}_D$ and their associated higher order perturbation terms are estimated in Appendix A and are of the order of $10^{-12} \sim 10^{-11}$ eV. Thus the S-T anticrossing gap is mainly determined by $H^{SO}_Z$ and can be made to be of the order of $10^{-8}$ eV by using the group IV semiconductors and by tuning the Rashba SO coupling constant. On the other hand, the hyperfine coupling energy in the localized electron system is about $10^{-7} \sim 10^{-6}$ eV as discussed in Sec. IV. Thus, in these systems our protocol of the nuclear spin quantum memory can be fully effective.

\section{Hyperfine interaction and quantum state transfer}

 We describe here the protocol for the quantum state transfer (QST) between a pair of electrons and a single nucleus. As such examples, there are a single electron charged $^{28}$Si QD with a single P atom, a two-electrons charged $^{28}$Si QD with a $^{29}$Si atom, a single-electron charged ZnSe QD with a single F atom, a two-electrons charged ZnSe QD with a $^{77}$Se isotope atom, and similar structures of the group IV elemental and II-VI compound semiconductors.

The hyperfine (hf) interaction between a pair of electrons and a
nuclear spin is given by the contact hf interaction \cite{Abragam96}
\begin{eqnarray}
V_{hf}=\frac{8\pi}{3} g_e\mu_B \; g_n\mu_n \sum_{i=1, 2} {\bf S}_i\cdot{\bf I} \; \delta({\bf
r}_i-{\bf R})\;,
\end{eqnarray}
where ${\bf S}({\bf I})$ is the electron (nuclear) spin operator in the dimensionless form, $\mu_B (\mu_n)$ the Bohr (nuclear) magneton, $g_e (g_n)$ the electron (nuclear) g-factor and ${\bf R}$ denotes the position of the nucleus.
For the QST protocol we will be interested in the singlet($S$) state and the
triplet($T_+$) state with the magnetic quantum number $m=1$ of an electron pair. 
 Here we consider for simplicity the case of two delocalized electrons in a QD containing a single isotope atom of the host material with a nuclear spin. The orbital function $\phi_g (\phi_e)$ is the ground (first excited) state with the magnetic quantum number m=0 (1) of the solutions of the effective mass equation Eq. (\ref{eme}):
\begin{eqnarray}
\phi_g=\frac{1}{\ell} \sqrt{\frac{2}{\pi d}} \; \exp\{-\frac{r^2}{2\ell^2}\}\cos{\frac{\pi z}{ d}} \;, \label{eqg} \\
\phi_e=\frac{1}{\ell} \sqrt{\frac{2}{\pi d}} \; \frac{r}{\ell}\exp\{-\frac{r^2}{2\ell^2}-i\varphi\}\cos{\frac{\pi z}{ d}} \;, \label{eqe}
\end{eqnarray}
where $\ell$ is the harmonic confinement length, $d$ the
thickness of QD and the $z$ coordinate lies in the range of $[-d/2,d/2]$. Then the wavefunctions of the singlet $S$ state and the triplet $T_+$ state are given by
\begin{eqnarray}
&&|S\rangle=\phi_g({\bf r_1}) \phi_g({\bf r_2}) \frac{1}{\sqrt{2}}(\alpha(\xi_1)\beta(\xi_2)
- \beta(\xi_1)\alpha(\xi_2)) \;,\\
&&|T_+\rangle=\frac{1}{\sqrt{2}}[\phi_g({\bf r_1}) \phi_e({\bf r_2})- \phi_e({\bf r_1}) 
\phi_g({\bf r_2})]  \; \alpha(\xi_1) \alpha(\xi_2) \;,
\end{eqnarray}
where $\alpha$ and $\beta$ are the spin up and down functions and $\xi$ denotes the spin coordinate.

In the vicinity of the S-T$_+$ crossing, the hf interaction comes
into play, inducing a flip-flop process between the electron spin pair and
the nuclear spin. The relevant Hamiltonian near the crossing point can be derived by calculating the matrix elements of the $V_{hf}$. It is to be noted that the actual wavefunction of an electron in solids is a product of the Bloch function $u({\bf r})$ and the envelope function $F({\bf r})$ in, e.g., Eqs. (\ref{eqg}) and (\ref{eqe}):
\begin{eqnarray}
&&\Psi({\bf r})= F({\bf r}) \;u({\bf r}) \\
{\rm with} && \frac{1}{v_0}\int_{v_0} d{\bf r} |u({\bf r})|^2 =1 \;, \; \int d{\bf r} |F({\bf r})|^2 =1 \;, 
\end{eqnarray}
where the Bloch function is normalized in the volume $v_0$ of a unit cell and is dimensionless and $F({\bf r})$ is normalized in the whole space. Now the relevant Hamiltonian is given as
\begin{eqnarray}
&&V_{hf}=h_{0} |T_+\rangle_e \;_e\langle T_+| \otimes (|\uparrow\rangle_n \;_n\langle\uparrow|- |\downarrow\rangle_n \;_n\langle\downarrow|) \nonumber \\
&& \quad + h_{1} (|T_+ \rangle_e |\downarrow\rangle_n \;_e\langle S| \;_n\langle \uparrow|+{\rm h.c.}) \label{stplus} \\
{\rm with} && h_0=\frac{{\cal A}}{4} \Bigl(|\phi_g({\bf R})|^2+|\phi_e({\bf R})|^2\Bigr) \;, \\
&&h_1=\frac{{\cal A}}{2} \phi_e^*({\bf R})\phi_g({\bf R}) \;, \label{h1eq} \\
&& {\cal A}=\frac{8\pi}{3} g_e \mu_B g_n \mu_n |u({\bf R})|^2 \;, \label{hfc}
\end{eqnarray}
where ${\bf R}$ denotes the 
site of the nuclear spin and the suffix $e(n)$ attached to the ket and bra vectors indicates the electron (nucleus). 

For the single electron charged QD doped with a donor atom, we can distinguish three regimes according to the donor position as discussed in Sec. III.  In the regime (c) where two electrons are delocalized throughout the QD, the hf coupling Hamiltonian is exactly the same as obtained above. In the regime (a) where two electrons are strongly bound by the donor atom, the same formulation as above can be used only by changing the meaning of the basis functions $\phi_g$ and $\phi_e$. Namely, $\phi_g$ and $\phi_e$ are the donor-bound ground and excited states, respectively. Then $h_1$ vanishes because $\phi_e$ is usually an odd-parity excited state which has a vanishing amplitude at the origin, i.e., at the donor nucleus.
On the other hand, in the regime (b) one electron is delocalized within the QD and the other electron is strongly bound by the donor atom. In this case the singlet and triplet states are given by
\begin{eqnarray}
&&|S\rangle=\frac{1}{2} (\phi_a({\bf r_1}) \phi_b({\bf r_2})+\phi_b({\bf r_1}) \phi_a({\bf r_2}))(\alpha(\xi_1)\beta(\xi_2)
- \beta(\xi_1)\alpha(\xi_2)) \;,\\
&&|T_+\rangle=\frac{1}{\sqrt{2}}[\phi_a({\bf r_1}) \phi_b({\bf r_2})- \phi_b({\bf r_1}) 
\phi_a({\bf r_2})]  \; \alpha(\xi_1) \alpha(\xi_2) \;,
\end{eqnarray}
where $\phi_a$ is the ground state in the QD confinement potential and $\phi_b$ is the ground state of the donor-bound electron. Then we find
\begin{equation}
h_1=\frac{\cal{A}}{2\sqrt{2}} (|\phi_b({\bf R})|^2- |\phi_a({\bf R})|^2) \;. \label{eqq2}
\end{equation}
Noting that in the regime (b) the donor atom is located off the center of the QD potential by a few times the confinement length $\ell$, we can neglect the second term in Eq. (\ref{eqq2}). The first term can be large because $\phi_b$ is a strongly localized function with a typical spatial extent of a few nm. For example, in a P-doped Si QD $h_1$ would be 41.4 MHz corresponding to the time constant of 24 ns, because ${\cal A}|\phi_b({\bf R})|^2$ is known to be 117 MHz from the experiments on Si:P samples\cite{Feher}. 
Consequently, the regime (b) of the single-electron charged QD doped with a donor atom is most favorable to realize a strong hf coupling and to achieve the electron-nuclear spin QST.

 Now we shall discuss the feasibility of the electron-nuclear spin QST for the two-electrons system. For the two-electrons charged ZnSe QD doped with an isotope $^{77}$Se atom having the nuclear spin($I$=1/2), the hf coupling constant would be about ${\cal A} = 3 \sim 4 \times 10^{-8}$ eV (nm$^3$), if $|u(R)|^2=100$ is assumed for the Bloch function referring to the value 186 in the case of Si\cite{Shulman}.
For a typical QD with $\ell \simeq 20$ nm and $d \simeq 10$ nm, the magnitude of the
hf interaction $h_1$ is estimated as 
\begin{equation}
h_1 = 2 \sim 3 \times (|{\bf R}|/\ell) \exp[-|{\bf R}|^2/\ell^2] \times 10^{-12}\; {\rm eV} \cong 2 \sim 3 \; \times 10^{-12}\; {\rm eV} \;,
\end{equation}
where the nuclear spin is assumed to be located at the midpoint $z=0$ in the z-direction and the factor related to the lateral part of wavefunctions is simply assumed to be 1. The corresponding electron-nuclear spin QST time is 1$\sim$ 2 ms. The singlet (S)-triplet (T) anticrossing gap due to the SO interaction should be much smaller than the hf coupling energy. As shown in Appendix A, the S-T anticrossing gap is determined by the matrix element $\langle S|V^{eff}_{SO}|T_{\pm} \rangle$, where $V^{eff}_{SO}$ is the effective SO interaction after a unitary transformation to eliminate the original SO interaction. The most dominant term is the SO induced Zeeman interaction $H^{SO}_Z$ and this term can be made to vanish in the first order by the magic angle tuning of the direction of the magnetic field. But the actual value of the S-T anticrossing gap is determined by the higher order perturbation terms concerning $H^{SO}_Z$ and contributions from the renormalized SO interaction $H_{ren}$. The typical magnitude of the S-T anticrossing gap is about $10^{-12} \sim 10^{-11}$ eV as shown in Appendix A. Thus the S-T anticrossing gap is of the same order of magnitude as the hf interaction energy.
On the other hand, in Si, the Dresselhaus SO terms are absent due to the inversion symmetry of the crystal lattice. According to the experimental report on a SiGe/Si/SiGe quantum well\cite{Wilamowski}, the Rashba SO coupling constant $\hbar a_R$ is about 0.55 $\times 10^{-4}$ eV \AA.  In this case also, the S-T anticrossing gap is estimated to be about $10^{-12} \sim 10^{-11}$ eV.
The hf coupling constant for the $\;^{28}$Si QD with a $\;^{29}$Si($I$=1/2) isotope atom is calculated as ${\cal A} = 5.06 \times 10^{-8}$ eV (nm$^3$), employing $|u(R)|^2=186$\cite{Shulman} and $h_1$ is estimated as 
\begin{equation}
h_1 = 4.0 \times 10^{-12} \; {\rm eV} 
\end{equation}
for $\ell = 20$ nm and $d = 10$ nm. 
Thus, in Si and ZnSe, the S-T anticrossing gap is of the same order of magnitude as the hf coupling energy. Consequently, our protocol for the electron-nuclear spin QST will not be effective.

On the other hand, in the single-electron charged QD with a donor impurity having the nuclear spin, the stronger hf coupling and the faster electron-nuclear spin QST can be expected because of the highly localized nature of the donor wavefunction. 
In the single-electron charged Si QD doped with a P donor, two electrons composed of the donor electron and an externally added electron play the role of a qubit. As discussed above, the hf coupling is the strongest for the regime (b) of Fig. \ref{donor}, in which one electron is delocalized within the QD and the other electron is strongly localized around the donor atom, and the magnitude would be 41.4 MHz. Concerning the spin-orbit interaction, the Dresselhaus terms are absent and the Rashba term gives rise to the S-T anticrossing gap of the order of $10^{-8}$ eV, as shown in Appendix A. For the single-electron charged ZnSe QD doped with a F donor, the situation is the same, although the S-T anticrossing gap would be larger. For this system the hf coupling constant would be a few tens of MHz according to the estimation in Appendix D.
 Thus, the single-electron charged QDs of the group IV elemental and II-VI semiconductors doped with a donor impurity is very favorable to realize the electron-nuclear spin QST and the nuclear spin quantum memory.

 In order to realize QST, a flip-flop type of interaction is essential and in a QD occupied by a pair of electrons this can be realized at the S-T$_+$ crossing. When the nuclear spin is
initialized in the $\downarrow$ state, the system evolves as
\begin{eqnarray}
&&\psi(t=0)=(a |S\rangle_e + b |T_+\rangle_e) \otimes |\downarrow\rangle_n \;, \\
&&\psi(t)= a |S\rangle_e |\downarrow\rangle_n+ b \Bigl(\cos \frac{h_1t}{\hbar} \;|T_+\rangle_e |\downarrow\rangle_n 
-i\sin \frac{h_1t}{\hbar} \;|S\rangle_e |\uparrow\rangle_n) \;,  \label{eqq4} \\
&&\psi(t=\frac{\pi\hbar}{2h_1})=|S\rangle_e \otimes ( a|\downarrow\rangle_n-i b |\uparrow\rangle_n) \:,
\end{eqnarray}
where $a$ and $b$ are arbitrary constants normalized as $|a|^2+|b|^2=1$.
Thus the quantum state of the electron pair is transferred to the nuclear spin. After that by tuning the magnetic field strength off the S-T$_+$ crossing point, the hf interaction is effectively switched off and the nuclear spin memory can be preserved. In the retrieval process, we prepare the singlet state of an electron pair in the QD and tune the magnetic field strength just at the S-T$_+$ crossing point. Then by waiting for a time $\pi \hbar/(2h_1)$, the state evolves as
\begin{equation}
\psi(t')=|S\rangle_e \otimes ( a|\downarrow\rangle_n-i b |\uparrow\rangle_n) \rightarrow \psi(t'+\frac{\pi \hbar}{2h_1}) =(a |S\rangle_e - b |T_+\rangle_e) \otimes |\downarrow\rangle_n \;.
\end{equation}
This state is not exactly the original state but the sign change of $b$ can be remedied by the optical STIRAP process which rotates the pseudospin spanned by the $|S\rangle$ and $|T_+\rangle$ states\cite{Takagahara09}. If we want to recover the original state only by the hf interaction, we have to wait for a time $3\pi \hbar/(2h_1)$. In the above we have neglected the nuclear Zeeman energy. When this Zeeman energy is taken into account, another phase factor is attached to $b$ in Eq. (\ref{eqq4}). This phase factor, however, can be cancelled by inverting the direction of the magnetic field in the retrieval stage. 

In this scheme, a knowledge of the hf interaction constant $h_1$ is necessary to achieve QST. 
 Now we discuss how the hf interaction constant $h_1$ can be determined.
 This is a sample specific information and can be fixed by measuring the probability of the spin state mixing between the singlet and triplet states of two electrons as a function of the hf interaction period, in a similar way to the experiment carried out by Petta et al.\cite{Petta05}. One has to initialize the two electrons in the singlet state, then to adjust the magnetic field strength to the singlet-triplet crossing point and to wait for a period $\tau$. Next, tuning away from the singlet-triplet crossing point, one measures the probability to find the electrons in the singlet state. The measurement is repeated at a long interval during which the nuclear spin is randomized. The probability to obtain the outcome of the singlet state should follow
\begin{equation}
 P(\tau)=\frac{1}{2} \Bigl[1+\cos^2(\frac{h_1\tau}{\hbar}) \Bigr] \;,
\end{equation}
from which the magnitude of the hf interaction constant $h_1$ can be estimated.

Nuclear spin initialization can as well be realized through the electron spin
measurements. Assume that a pair of electrons is initialized in the spin singlet state,
whereas the nuclear spin is in an arbitrary mixed state, then the system is
tuned to the S-T$_+$ crossing point for a time $\tau$. The time evolution of the system is given by
\begin{eqnarray}
&&\rho(t=0)=|S\rangle_e \;_e\langle S| \otimes \Bigl(p_{\uparrow} \;|\uparrow\rangle_n \;_n\langle\uparrow|
+p_{\downarrow}\; |\downarrow\rangle_n \;_n\langle\downarrow| \Bigr) \;, \label{init}\\
&&\rho(\tau)=p_{\uparrow} \;\mat{\psi(\tau)}{\psi(\tau)}+p_{\downarrow} \;|S\rangle_e \;_e\langle S| \otimes |\downarrow\rangle_n \;_n\langle \downarrow|  \\
{\rm with} &&\psi(\tau)=\cos \frac{h_1\tau}{\hbar} \;|S\rangle_e |\uparrow\rangle_n -i\sin \frac{h_1\tau}{\hbar}\;|T_+\rangle_e |\downarrow\rangle_n.
\end{eqnarray}
At the next step an electron spin measurement is carried out in the
singlet-triplet basis. The probability $P_S$ to have the outcome of the singlet state and the density matrix $\rho_S$ after the measurement are given by
\begin{eqnarray}
&&P_S=p_{\downarrow}+ p_{\uparrow} \cos^2 \frac{h_1\tau}{\hbar} \;,\\ 
&&\rho_S=\frac{1}{P_S} |S\rangle_e \;_e\langle S| \otimes \Bigl(p_{\downarrow} \;|\downarrow\rangle_n \;_n\langle\downarrow|+p_{\uparrow} \;\cos^2 \frac{h_1\tau}{\hbar} \; |\uparrow\rangle_n \;_n\langle\uparrow| \Bigr) \;. \label{singlet}
\end{eqnarray}
In the same way, for the outcome of the triplet state the corresponding quantities are given by
\begin{equation}
P_T=p_\uparrow\sin^2 \frac{h_1\tau}{\hbar} \;,\quad \rho_T=|T_+\rangle_e \;_e\langle T_+| \otimes |\downarrow\rangle_n \;_n\langle\downarrow| \;.
\end{equation}
If one knows the hf interaction constant, then setting
$\tau=\pi \hbar/(2h_1)$ guarantees that the nuclear spin is initialized in the $\downarrow$ state. Otherwise, i.e., when one doesn't
know the exact hf interaction constant, the initialization goes as follows: if the triplet state is detected, the nuclear spin state is initialized in the $\downarrow$ state
deterministically. If the singlet state is detected, in the
conditional state $\rho_S$(Eq. (\ref{singlet})), the weight of
the $\uparrow$ nuclear spin state is decreased with respect to the
initial state $\rho(t=0)$(Eq. (\ref{init})).  Thus by continuing the electron spin
measurement, one can eventually purify the nuclear spin state into the
$\downarrow$ spin state.

\section{Adiabatic quantum state transfer}

\begin{figure}
\includegraphics{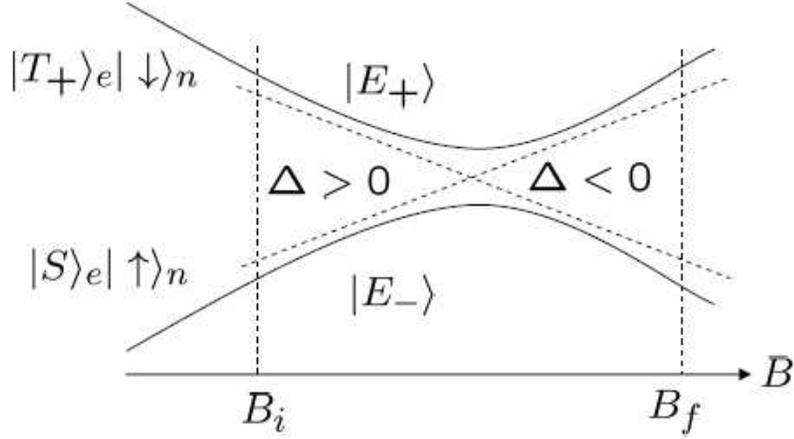}
\caption{Schematic energy level diagram around the $S-T_+$ crossing point. Dashed lines ($\dn{T_+}_e \dn{\downarrow}_n$ and $\dn{S}_e \dn{\uparrow}_n$) represent the original energy levels without the hf coupling, whereas solid lines ($\dn{E_+}$ and $\dn{E_-}$) depict the energy levels including the hf coupling. $\Delta$ denotes the energy difference between $\dn{T_+}_e\dn{\downarrow}_n$ and $\dn{S}_e\dn{\uparrow}_n$. The adiabatic state transfer is achieved by sweeping the magnetic field from $B_i$ to $B_f$ and backward.} \label{cross}
\end{figure}

In the last Section, the scheme of the quantum state transfer between two electrons and a nuclear spin is presented under the assumption that the hf coupling constant is known. However, there is an alternative scheme to do the same quantum state transfer without the knowledge of the hf coupling constant. This is based on the adiabatic state control by the magnetic field tuning through the S-T$_+$ crossing point, which is schematically illustrated in Fig. \ref{cross}. 

The relevant Hamiltonian is given by
\begin{eqnarray}
&&H=\frac{-\Delta}{2}\;|S\rangle_e \;_e\langle S| \; (|\uparrow\rangle_n \;_n\langle\uparrow|+ 
|\downarrow\rangle_n \;_n\langle\downarrow|)+ \frac{\Delta}{2} |T_+\rangle_e \;_e \langle T_+| 
|\downarrow\rangle_n \;_n\langle\downarrow| \nonumber \\
&& +h_1\; (|T_+\rangle_e |\downarrow\rangle_n \;_e\langle S| \;_n\langle\uparrow|+|S\rangle_e |\uparrow\rangle_n \;_e\langle T_+| \;_n\langle\downarrow|) \;, \label{ham1}
\end{eqnarray}
where $\Delta$ is the energy splitting between the singlet and triplet states, $h_1$ the hf interaction constant derived in Eq. (\ref{h1eq}) and the nuclear Zeeman energy is omitted.
The electron spin states $|T_+\rangle_e$ and $|S\rangle_e$ are coupled by the hyperfine interaction, resulting in the eigenstates:
\begin{eqnarray}
&&\dn{E_\pm}=\frac{1}{\sqrt{N_\pm}}\bigl[ -h_1\dn{T_+}_e \dn{\downarrow}_n+(\Delta/2\mp E_0)\dn{S}_e \dn{\uparrow}_n\bigr] \\
{\rm with} \; && N_\pm=h_1^2+(\Delta/2\mp E_0)^2 
\end{eqnarray}
and the energies
\begin{equation}
E_\pm=\pm E_0 \;, \; E_0=\sqrt{\Delta^2/4+h_1^2} \;.
\end{equation}

Initially an arbitrary electron state is prepared with the nuclear spin in the $\dn{\downarrow}_n$ state. Then the magnetic field is sweeped to the right from $B=B_i$ at $t=t_i$ to $B=B_f$ at $t=t_f$, as shown in Fig. 5 adiabatically and the parameter $\Delta$ is tuned from an initially positive value $\Delta_i$ to a negative value $\Delta_f$, where $|\Delta_{i, f}| \gg h_1$ is assumed. At $B=B_i$ or $\Delta=\Delta_i$ the state $|E_+\rangle$ is composed mostly of $\dn{T_+}_e \dn{\downarrow}_n$, whereas at $B=B_f$ or $\Delta=\Delta_f$ the state $|E_+\rangle$ is composed mostly of $\dn{S}_e \dn{\uparrow}_n$. In the course of this sweeping, the $\dn{T_+}_e \dn{\downarrow}_n$ state starts from the $|E_+\rangle$ branch in Fig. 5 and continues to be on the same branch but an extra phase factor is acquired. The time evolution of the wavefunction is given by 
\begin{eqnarray}
&&\dn{\psi_0}=[ a \dn{S}_e+ b \dn{T_+}_e] \dn{\downarrow}_n\rightarrow \dn{\psi_1}=\dn{S}_e \;[\;a \dn{\downarrow}_n e^{i\phi_1}+b \dn{\uparrow}_n e^{i\phi_2}]  \label{psieq} \\
{\rm with} && \phi_1=\frac{1}{2 \hbar} \int_{t_i}^{t_f}d t \,\Delta(B) = \frac{1}{2 \hbar} \int_{B_i}^{B_f}d B \frac{1}{|\frac{d B}{d t}|} \; \Delta(B) \;, \\
&&\phi_2=-\frac{1}{\hbar} \int_{t_i}^{t_f} d t \; \sqrt{\Delta^2(B)/4+h_1^2} = -\frac{1}{\hbar} \int_{B_i}^{B_f} d B \frac{1}{|\frac{d B}{d t}|} \; \sqrt{\Delta^2(B)/4+h_1^2} \;,
\end{eqnarray}
where $\Delta$ is a function of the magnetic field $B$, $B$ is a function of the time $t$ and $h_1$ is assumed to be constant. As a consequence, an additional phase difference $\phi_2-\phi_1$ is introduced in the superposition state of the nuclear spin. However, this additional phase can be eliminated at the retrieval stage. A radio frequency $\pi$ pulse is applied on the state $|\psi_1 \rangle$, swapping the nuclear spin states $\dn{\uparrow}_n \leftrightarrow \dn{\downarrow}_n$ and the state becomes
\begin{equation}
\dn{\psi_2}=\exp[iI_x\pi]\dn{\psi_1}=\dn{S}_e \;[\;a \dn{\uparrow}_n e^{i\phi_1}+ b \dn{\downarrow}_n e^{i\phi_2}] \;,
\end{equation}
where $I_x$ is the $x$-component of the nuclear spin operator ($|I|=1/2$).

Then the adiabatic sweeping is reversed from $B_f$ at $t=t'_i$ to $B_i$ at $t=t'_f$ with the same time profile. In this case, the state $\dn{S}_e \dn{\uparrow}_n$ transfers adiabatically to the state $\dn{T_+}_e \dn{\downarrow}_n$, although another extra phase factor is acquired:
\begin{equation}
\phi'_2=-\frac{1}{\hbar} \int_{t'_i}^{t'_f} d t \; \sqrt{\Delta^2(B)/4+h_1^2}
=-\frac{1}{\hbar} \int_{B_i}^{B_f} d B \frac{1}{|\frac{d B}{d t}|} \; \sqrt{\Delta^2(B)/4+h_1^2} =\phi_2 \;.
\end{equation}
In the same way, the state $\dn{S}_e \dn{\downarrow}_n$ acquires the phase shift
\begin{equation}
\phi'_1=\frac{1}{2 \hbar} \int_{t'_i}^{t'_f} d t \,\Delta(B) = \frac{1}{2\hbar} \int_{B_i}^{B_f} d B \frac{1}{|\frac{d B}{d t}|} \; \Delta(B) =\phi_1 \;.
\end{equation}
Consequently, the wavefunction becomes
\begin{equation}
\dn{\psi_2} \rightarrow a \dn{T_+}_e \dn{\downarrow}_n e^{i\phi_1+i\phi'_2}+ b \dn{S}_e \dn{\downarrow}_n  e^{i\phi_2+i\phi'_1} 
=[ a \dn{T_+}_e + b \dn{S}_e] \dn{\downarrow}_n e^{i\phi_1+i\phi_2} \;.
\end{equation}
This state is not just the original state $\dn{\psi_0}$ in Eq. (\ref{psieq}) but with a swap between the spin singlet and triplet states.
However, the above protocol can be repeated once more to obtain the original electron spin state. Alternatively, we can employ an optical STIRAP (stimulated Raman adiabatic passage) method to perform the rotation within the pseudospin subspace composed of the states $|S\rangle_e$ and $|T_+\rangle_e$\cite{Takagahara09}. Then the original state is recovered.

 Finally, the condition for the adiabatic state transfer will be discussed. When the magnetic field is sweeped, the parameter $\Delta$ in the Hamiltonian (\ref{ham1}) is time-dependent. The wavefunction in the subspace spanned by $\dn{S}_e \dn{\uparrow}_n$ and $\dn{T_+}_e \dn{\downarrow}_n$ can be expanded as
\begin{equation}
\dn{\psi(t)}=a(t) \dn{E_+} + b(t) \dn{E_-} \;,
\end{equation}
where the eigenstates $\dn{E_+}$ and $\dn{E_-}$ are also time-dependent through the time-dependence of the parameter $\Delta$. By noting that
\begin{eqnarray}
&&\frac{d}{d t} \dn{E_+}=\frac{h^2_1 \dot{\Delta}}{N^{1/2}_- \;N^{3/2}_+} (E_0- \Delta/2) \dn{E_-}=c_+(t) \; \dn{E_-} \;, \\
&&\frac{d}{d t} \dn{E_-}=- \frac{h^2_1 \dot{\Delta}}{N^{1/2}_+ \;N^{3/2}_-} (E_0+ \Delta/2) \dn{E_+}=c_-(t) \; \dn{E_+} \;,
\end{eqnarray}
where $c_+(t)$ and $c_-(t)$ are introduced for simplicity, we have the Schr\"odinger equations:
\begin{eqnarray}
&& \frac{d}{d t} a(t)= -\frac{i}{\hbar} E_+(t) \; a(t) +c_-(t) \; b(t) \;,\\
&& \frac{d}{d t} b(t)= -\frac{i}{\hbar} E_-(t) \; b(t) +c_+(t) \; a(t)\;.
\end{eqnarray}
When the second terms on the right hand side of the above equations can be neglected, the adiabatic time-evolution is guaranteed. This condition will be examined near the S-T$_+$ crossing point where $\Delta \sim 0$ and the adiabaticity condition is most stringent because the $\dn{E_{\pm}}$ branches are close to each other. Around the S-T$_+$ crossing point, $E_+ \cong h_1\;, \; E_- \cong - h_1$ and $c_+(t)$ and $c_-(t)$ are estimated as
\begin{equation}
c_+(t) \cong - c_-(t) \cong \frac{\dot{\Delta}}{4 h_1}=c_0 \;. 
\end{equation}
Then the time evolution under the initial condition $a(0)=1$ and $b(t)=0$ is given by
\begin{eqnarray}
&&a(t)=\cos \Omega t -i \frac{\omega_1}{\Omega} \sin \Omega t \;, \; b(t)=\frac{c_0}{\Omega} \sin \Omega t \\
{\rm with} && \omega_1=\frac{h_1}{\hbar} \;, \; \Omega=\sqrt{\omega^2_1 + c^2_0} \;.
\end{eqnarray}
The amplitude $b(t)$ should be small enough to guarantee the adiabaticity, namely, 
\begin{equation}
|\frac{\dot{\Delta}}{4 h_1}| \ll |\frac{h_1}{\hbar}| \longrightarrow |\dot{\Delta}|=|\frac{d B}{d t} \frac{d \Delta}{d B}| \ll \frac{4 h^2_1}{\hbar} \:.
\end{equation}
This condition restricts the speed of the magnetic field sweeping for the adiabatic state transfer.

\section{Quantum state transfer between a single electron and a single nuclear spin}

So far we have considered the qubit composed of a pair of electrons. However, the single electron qubit is more fundamental as a building block for the quantum information processing.
Here we consider a system composed of a single electron as a qubit and a nuclear spin as a quantum memory whose examples will be discussed later.
It is possible to devise a scheme for the quantum state transfer between a single electron spin and a nuclear spin.
 At zero magnetic field the hyperfine interaction between a single electron spin and a nuclear spin is given by
\begin{equation}
H=A \; {\bf I}\cdot {\bf S} \;, \quad A=\frac{8\pi}{3} g_e \mu_B g_n \mu_n \; |F({\bf R}) u({\bf R})|^2 \;,
 \label{ish}
\end{equation}
where ${\bf I}({\bf S})$ denotes the spin-1/2 nuclear (electronic) spin, $u({\bf r})(F({\bf r}))$ is the Bloch (envelope) function of the electron and ${\bf R}$ is the position vector of the nucleus. 
In the hf interaction the dipolar term vanishes, since the dynamics takes place within the electronic ground state orbital \cite{Abragam96} and the Fermi contact interaction is dominant. A magnetic field is applied in order to switch off the hf coupling so that the large difference in the Zeeman energy splitting between an electron and a nucleus prohibits the flip-flop transitions between them.
As shown in the Appendix C, under the Hamiltonian in Eq. (\ref{ish}), the time-evolution of the electron-nucleus coupled system proceeds as follows:
\begin{eqnarray}
&&\rho\;(t=0)=|\psi\rangle_e \;_e\langle\psi| \otimes \rho_n 
\longrightarrow  \rho\;(t=\pi \hbar/A)= \rho_e \otimes |\psi\rangle_n \;_n\langle\psi| \\
{\rm with} && |\psi\rangle= \alpha|\uparrow\rangle+\beta|\downarrow\rangle\;,
\end{eqnarray}
where $\alpha$ and $\beta$ are arbitrary complex constants normalized as $|\alpha|^2+|\beta|^2=1$, the suffix $e(n)$ indicates the electron (nucleus), $\rho_n$ is a density matrix representing an arbitrary mixed state of the nucleus and $\rho_e$ is the same mixed state for the electron. 
Thus the quantum state transfer is possible for an arbitrary mixed state of the nuclear spin and the initialization of the nuclear spin is not necessary.

In the first stage of the QST, after the hf interaction of duration $\tau=\pi\hbar/A$ the spin states are swapped between the electron and the nucleus. Then a magnetic field is applied to switch off the hf coupling and to preserve the nuclear spin memory. After some interval within the nuclear spin coherence time, the magnetic field is turned off and the hf interaction is again switched on for another time period of $\tau=\pi \hbar/A$. Then the initial quantum state is retrieved back in the electron spin. Actually, the magnetic field applied during the storage period introduces an extra phase factor due to the nuclear Zeeman energy splitting. But this phase can be cancelled by applying a magnetic field after retrieving the quantum state in the electron spin. 

This scheme of QST is relevant for a single electron trapped in a single QD and also for a neutral donor in the bulk semiconductor which is composed of isotope atoms without the nuclear spin. For instance, the Si:P system has been extensively studied and has already been proposed as a qubit for the scalable quantum computer\cite{Kane98,Skinner,Vrijen}. $^{31}$P has a nuclear spin $I=1/2$ with the hf interaction constant of $A=117$ MHz. Thus the QST can be realized on a time scale about $10$ ns, which is much shorter than the electron coherence time which can extend to about 60 ms \cite{Tyryshkin03,Lyon08}.
Another candidate might be a II-VI semiconductor because in this material the abundance of isotope atoms with the nuclear spin is only a few percent and the isotope purification may be possible. For example, ZnSe has the natural abundance of $4.1 (7.6) \%$ for the $^{67}$Zn ($^{77}$Se) isotope atom with the nuclear spin $I=5/2 (1/2)$. Thus the isotopically purified and nuclear spin-free ZnSe crystal doped with a fluorine atom, namely ZnSe:F, would be a good 
example \cite{Sana}. So far, we are not aware of any NMR data for the ZnSe:F system. But we can
estimate the hyperfine coupling constant for the ZnSe:F by comparison with the case of Si:P, as shown in Appendix D. According to that, we infer that $A$(ZnSe:P) $\simeq$ 78.6 MHz, which indicates a smaller hf coupling compared with the Si:P case.

The important feature of these localized electron system is the optical interface. Because of the localized nature the optical transition is possible even in the indirect-gap materials like Si. Actually the photoluminescence from Si:P centers has been extensively studied \cite{Haynes,Dean}. The photoluminescence occurs through the neutral donor-bound exciton (D$^0$X) state which is composed of two electrons and one hole. This photoluminescence can be utilized to locate and address each localized center. In view of the recent progress in manipulating optically a single electron spin trapped in a QD \cite{Berezovsky,Press}, we can expect that 
the donor electron spin can also be controlled optically via the $\Lambda$-type transition through the donor-bound exciton state. Another important feature of the localized electron system is the uniformity of the system. For example, any Si:P center has the same optical transitions like atoms because the nearby atomic configuration is the same for any center. Thus a laser light of the same wavelength can be used for the optical initialization, manipulation and measurement of the electron spin in any localized center. This feature is advantageous in constructing a scalable quantum network from an array of these localized centers through the optical interconnection.

\begin{table}
\caption{Properties related to the quantum state transfer (QST) for the two-electron and single-electron qubits. Schematic configurations of electrons and an impurity for cases (a) to (f) are illustrated in Fig. 7.  \label{table}}
\begin{tabular}{|p{3cm}|p{2.1cm}|p{2.3cm}|p{2.1cm}|p{2.0cm}|p{2.0cm}|p{2.3cm}|}
\cline{2-7}
 \multicolumn{1}{c|}{} &\multicolumn{3}{|c|}{two-electron qubit}&\multicolumn{3}{|c|}{single-electron qubit}\\ 
\cline{2-7}
\multicolumn{1}{c|}{}& (a) Si QD & (b) ZnSe QD & (c) Si:P QD & (d) Si:P & (e) ZnSe:F & (f) ZnSe QD \\ \hline
Nuclear spin \newline initialization & necessary & necessary & necessary &  unnecessary & unnecessary & unnecessary \\ 
\hline 
Electron-nuclear QST time & $\sim$ 1 ms & $\sim$ 1 ms & $\sim$ 20 ns & $\sim$10 ns & $\sim$ 10 ns & $\sim$ 1 ms \\
\hline 
Spin-orbit vs. hyperfine coupling &  $ V_{SO} \sim V_{hf}$ & $V_{SO} \sim V_{hf}$ &  $ V_{SO} \ll V_{hf}$ & irrelevant & irrelevant & irrelevant \\
\hline
S-T$_{\pm}$ crossing & unfavorable & unfavorable & favorable & irrelevant & irrelevant & irrelevant \\
\hline
Optical Interface & hard & possible & good & good & good & possible \\
\hline
\end{tabular}

\end{table}

\begin{figure}
\includegraphics[width=15cm,clip]{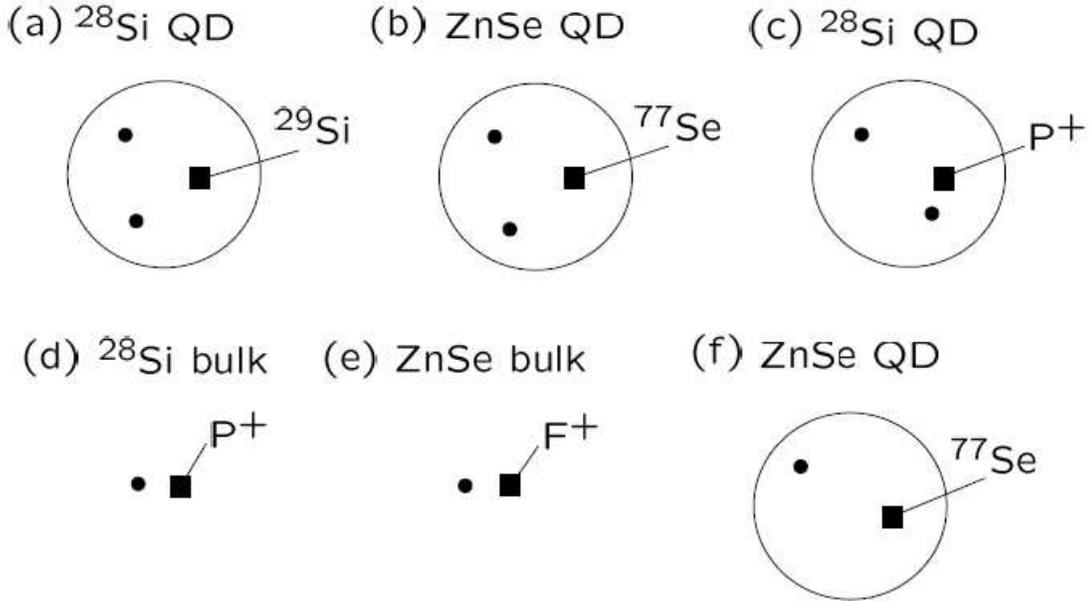}
\caption{Schematic configurations of electrons and an impurity for the electron-nuclear spin QST which are compared in Table I. The solid circle (square) represents an electron (impurity). The cases of (a), (b) and (c) correspond to the two-electron qubit, whereas the cases of (d), (e) and (f) to the single-electron qubit.}
\end{figure}

\section{Summary and Discussion}

We have proposed a few new protocols for the nuclear spin quantum memory in the isotopically purified group IV elemental and II-VI compound semiconductors, where the number of atoms with the nuclear spin is reduced to only one and thus the decoherence due to the nuclear dipole-dipole interaction can be avoided. We have studied various cases where the qubit is defined by the singlet and one of the triplet states of a two-electron system or simply by a single electron spin. For the two-electron system, we studied the case where two electrons are loaded on a donor-free QD with an isotope atom having the nuclear spin and also the case where one electron is additionally loaded on a QD doped with a single donor atom. In the latter case the behavior of the singlet-triplet crossing depends sensitively on the donor position, reflecting the hybridization between the donor-bound localized orbitals and the delocalized QD orbitals. Our protocol requires that the hyperfine coupling energy should be much larger than the singlet-triplet anticrossing gap. This requirement is satisfied favorably in the case of a donor doped QD because the large hyperfine coupling energy can be achieved only for strongly localized electrons.
In the two-electron system the 100\% nuclear spin initialization is possible based on the spin state measurement of electrons and can be utilized in the QST between the electronic and nuclear spins. Our protocol of the nuclear spin quantum memory can be achieved by the electrical control even
without tuning the magnetic field and complements the advent of the electrical gate tuning of
the singlet and triplet states of a pair of electrons \cite{Petta05}.

For the single electron system, we considered the case of a single donor electron coupled with the nuclear spin of the donor atom in the bulk crystal or in a QD. Here the QD is not essential but is convenient only for addressing each electron locally, because the donor electron is much more tightly bound to the donor atom compared with the delocalized electron in a QD. The most important feature in the single electron system is that the nuclear spin initialization is not required in the QST between the electronic and nuclear spins.
Typical examples of the single electron system are Si:P and ZnSe:F. These systems have the hf coupling constant about a hundred MHz and the QST time between the electron and the nucleus is about 10 ns. This feature is favorable in view of the long coherence times of the electron spin and the nuclear spin in the group IV elemental and II-VI compound semiconductors. Another example is the case of a single delocalized electron in a QD containing a single isotope atom with the nuclear spin of the host material, e.g., an isotopically purified $^{28}$Si QD with a $^{29}$Si atom and an isotopically purified ZnSe QD with a $^{77}$Se atom. Although the hf coupling constant for some cases is not well known, the constant can be inferred by assuming that $|u(R)|^2 \sim 100$ in Eq. (\ref{hfc}). According to this rough estimate, the hf coupling constant for a delocalized electron in a QD with a typical size of 20 nm is about a few times $10^{-12}$ eV and the QST time is about 1 ms. Thus the delocalized electron in a QD is not favorable compared with the donor-bound localized electron for the application to the nuclear spin quantum memory system. Furthermore, the donor-bound localized electron has, in general, a good optical interface and the optical initialization, manipulation and measurement of the electron spin would be possible through the donor-bound exciton state, although the experimental demonstration is yet to be challenged.

In Table \ref{table}, merits and demerits of the two-electron qubit and single-electron qubit are compared with respect to the time required for the QST between the electron and nuclear spins, the necessity of the nuclear spin initialization in the QST, the relative magnitude between the spin-orbit coupling and the hf interaction which determines the feasibility of our QST protocol, the possibility of the singlet-triplet state crossing for the two-electron qubit, and the feasibility of the optical interface. From this Table, we see that the donor-bound localized electron system is favorable with respect to the fast QST between the electron and nuclear spins, the irrelevance of the nuclear spin initialization in the QST and the feasibility of ultrafast optical manipulation and measurement of the electron spin. Furthermore, the localized electron system is homogeneous like atoms in the sense that the energy level structures and the associated optical transitions are the same for any localized center because of the characteristic nearby atomic configuration of a specific localized center. Based on these features, we envisage that an array of the donor-bound localized electrons provide an excellent set of qubits and the gate operation or entanglement transfer between any two localized qubits can be carried out through optical channels, as demonstrated recently using atomic qubits \cite{Olmschenk}, and consequently the quantum network can be established.

\begin{acknowledgments}
We would like to thank Professor H. Kosaka for stimulating discussions and continual encouragements. This work is financially supported by the Japan Science and Technology Agency and also by the Ministry of Education, Culture, Sports, Science and Technology. We also acknowledge the Supercomputer Center, Institute for Solid State Physics, University of Tokyo for the facilities and the use of the Hitachi SR11000.
\end{acknowledgments}

\newpage

\appendix
\section{ Spin-orbit interaction for a pair of electrons}

 The spin-orbit (SO) interaction for the conduction band electron in the linear approximation with respect to the momentum operator is given by
\begin{equation}
 V_{SO}=a_R (\sigma_x p_y-\sigma_y p_x)+ a_D (\sigma_x p_x-\sigma_y p_y) \label{Aso}
\end{equation}
in the two-dimensional limit, where the first term is the Rashba term due to the structural inversion asymmetry and the second term is the Dresselhaus term arising from the bulk inversion asymmetry. 
In the case of a pair of electrons it is convenient to rewrite the Hamiltonian in terms of the center-of-mass coordinate ${\bf R}=(X, Y, Z)$ and the relative coordinate ${\bf r}=(x, y, z)$, defined by
\begin{equation}
{\bf R}=\frac{1}{2}({\bf r_1}+ {\bf r_2}) \;, \; {\bf r}={\bf r_1}- {\bf r_2} \;, \label{aeqq2}
\end{equation}
where the subscript 1(2) refers to the first (second) electron.
Then we have
\begin{equation}
\frac{\p}{\p {\bf R}}= \frac{\p}{\p {\bf r_1}} + \frac{\p}{\p {\bf r_2}} \;, \; \frac{\p}{\p {\bf r}}= \frac{1}{2}(\frac{\p}{\p {\bf r_1}} - \frac{\p}{\p {\bf r_2}})
\end{equation}
and it is natural to introduce the momentum operators corresponding to the center-of-mass coordinate and the relative coordinate by
\begin{equation}
{\bf \Pi}={\bf p_1}+ {\bf p_2}\;, \; {\bf p}=\frac{1}{2}({\bf p_1} - {\bf p_2})\;. \label{mom}
\end{equation}
Under a uniform magnetic field the momentum operators are modified into the gauge-invariant form:
\begin{eqnarray}
&&{\bf p_i} = -i\hbar \frac{\p}{\p {\bf r_i}}+\frac{e}{c}{\bf A(r_i)}= -i\hbar \frac{\p}{\p {\bf r_i}}+\frac{e}{2c}{\bf B\times r_i} \quad (i=1, 2) \;, \label{aeq11} \\
&&{\bf \Pi} = -i\hbar \frac{\p}{\p {\bf R}}+ \frac{e}{c} {\bf B \times R} \;, \\
&&{\bf p} = -i\hbar \frac{\p}{\p {\bf r}}+ \frac{e}{4c} {\bf B \times r} \;,
\end{eqnarray}
where the symmetric gauge (${\bf A(r)=(B \times r)/2}$) is employed. Accordingly, the orbital part of the Hamiltonian is rewritten as
\begin{equation}
H=\frac{1}{4m}(\Pi^2_x +\Pi^2_y + \Pi^2_z)+ U_{cm}(X, Y, Z)+ \frac{1}{m}(p^2_x +p^2_y+p^2_z) +U_{rel}(x, y, z)+\frac{e^2}{\epsilon |{\bf r}|}\;, \label{ham}
\end{equation}
where $U_{cm}(U_{rel})$ are the circularly symmetric confinement potential for the center-of-mass (relative) coordinates and it is to be noted that the mass of the center-of-mass (relative) coordinate is $2m(m/2)$. 

Now the spin-orbit interaction in Eq. (\ref{Aso}) and the Zeeman energy $H_Z$ for two electrons can be written as
\begin{eqnarray}
 V_{SO}&&=\frac{1}{2} a_R (\Sigma_x\Pi_y - \Sigma_y\Pi_x) + a_R (\sigma_x p_y - \sigma_y p_x) \nonumber \\
 &&+\frac{1}{2} a_D (\Sigma_x\Pi_x - \Sigma_y\Pi_y) + a_D (\sigma_x p_x - \sigma_y p_y) \;,  \label{aeq1} \\
 H_Z&&=\frac{1}{2}g_e \mu_B {\bf B \cdot \Sigma} \\
 {\rm with} &&\Sigma_i=\sigma_{1i}+\sigma_{2i} \;, \; \sigma_i=\sigma_{1i}-\sigma_{2i}\quad (i=x, y, z) \;,
 \label{aeq2}
 \end{eqnarray}
where ${\bf \Pi}$ and ${\bf p}$ are defined in Eq. (\ref{mom}). Now we examine the effect of this spin-orbit interaction on the singlet-triplet level crossing. Since the Hamiltonian in Eq. (\ref{ham}) is separated into the center-of-mass and the relative coordinates, the singlet and triplet eigenstates can be written as
\begin{eqnarray}
&&|S\rangle=g_{cm}({\bf R})g_{rel}({\bf r}) \frac{1}{\sqrt{2}}(\alpha(\xi_1)\beta(\xi_2)- \beta(\xi_1)\alpha(\xi_2)) \;,\\
&&|T_+(T_-)\rangle=g_{cm}({\bf R})e_{rel}({\bf r}) \; \alpha(\xi_1)\alpha(\xi_2) \;(\beta(\xi_1)\beta(\xi_2)) \;,\\
&&|T_0\rangle=g_{cm}({\bf R})e_{rel}({\bf r}) \frac{1}{\sqrt{2}}(\alpha(\xi_1)\beta(\xi_2)+ \beta(\xi_1)\alpha(\xi_2)) \;,
\end{eqnarray}
where $g_{cm}(g_{rel})$ is the ground state of the center-of-mass (relative) motion, $e_{rel}$ is an odd-parity excited state of the relative motion, $\alpha(\xi)$ and $\beta(\xi)$ are the spin up and down functions and $\xi$ denotes the spin coordinate. The total spin operators $\Sigma_x$ and $\Sigma_y$ do not change the magnitude of the total spin but they and the difference spin operators $\sigma_x$ and $\sigma_y$ change the magnetic quantum number only by $\pm$1. Thus the following matrix elements vanish:
\begin{eqnarray}
&&\langle S|V_{SO}|S \rangle= \langle T_{\pm}|V_{SO}|T_{\pm} \rangle= \langle T_{\pm}|V_{SO}|T_{\mp} \rangle= \langle S|V_{SO}|T_0 \rangle=\langle T_0|V_{SO}|T_0 \rangle=0 \;.
\end{eqnarray}
In the matrix element $\langle S|V_{SO}|T_{\pm} \rangle$, $\Sigma_x$ and $\Sigma_y$ do not contribute because they do not change the magnitude of the total spin and thus only the $\sigma_x$ and $\sigma_y$ contribute. After the spin part is calculated, we find
\begin{equation}
\langle S|V_{SO}|T_{\pm} \rangle=\langle g_{cm} g_{rel}|i\sqrt{2} a_R (p_x\pm ip_y)\mp \sqrt{2} a_D (p_x\mp ip_y) |g_{cm} e_{rel}\rangle.
\end{equation}
Then, using the relations: 
\begin{equation}
[x, H]=\frac{i\hbar}{m/2}\; p_x \;, \; [y, H]=\frac{i\hbar}{m/2}\; p_y \;,
\end{equation}
we have
\begin{eqnarray}
&&\langle g_{cm} g_{rel}|p_x\pm i p_y|g_{cm} e_{rel}\rangle=\frac{m}{2 i\hbar} \langle g_{cm} g_{rel}|[x\pm iy, H]|g_{cm} e_{rel}\rangle  \nonumber \\
&&=\frac{m \Delta_{ST}}{2i\hbar} \langle g_{rel}|x\pm iy|e_{rel}\rangle=\frac{m \Delta_{ST}}{2i\hbar} \langle g_{rel}|r e^{\pm i\varphi}|e_{rel}\rangle \\
{\rm with}\quad && \Delta_{ST}=E_Z(S) -E_Z(T_{\pm}) \;,
\end{eqnarray}
where $E_Z(\lambda)$ denotes the Zeeman energy of the $\lambda$ state. Finally, the matrix element is calculated as
\begin{equation}
\langle S|V_{SO}|T_{\pm} \rangle=\frac{m \Delta_{ST}}{\sqrt{2}\hbar} [a_R \langle g_{rel}|r e^{\pm i\varphi}|e_{rel}\rangle \pm i a_D \langle g_{rel}|r e^{\mp i\varphi}|e_{rel}\rangle ] \:. \label{aeq6}
\end{equation}
This matrix element is finite in general because the S-T$_{\pm}$ crossing occurs at a finite magnetic field.

 However, we have to consider the higher order perturbation terms with respect to the 
spin-orbit interaction $V_{SO}$. For that purpose it is convenient to apply a unitary transformation to the single electron Hamiltonian:
\begin{equation}
H=\frac{1}{2m} {\bf p}^2+ a_R (\sigma_x p_y-\sigma_y p_x)+ a_D (\sigma_x p_x-\sigma_y p_y) +\frac{1}{2}g_e \mu_B \bm{ B}\cdot \bm{\sigma} \;,
\end{equation}
where ${\bf p}$ is the kinetic momentum vector defined by Eq. (\ref{aeq11}).
By rotating the coordinate system by an angle $\pi/4$ in the $xy$ plane, namely, by introducing a new coordinate system ($\xi, \eta, z$) defined by ${\bf e}_{\xi}=(1,1,0)/\sqrt{2}$, ${\bf e}_{\eta}=(-1,1,0)/\sqrt{2}$, ${\bf e}_z=(0,0,1)$, the relevant vector components are transformed as
\begin{eqnarray}
&&\left(\begin{array}{c} x \\ y \end{array}\right) =\frac{1}{\sqrt{2}} \left(\begin{array}{cc} 1 & -1 \\ 1 & 1 \end{array}\right) \left(\begin{array}{c} \xi \\ \eta \end{array}\right) \;, \;
\left(\begin{array}{c} p_x \\ p_y \end{array}\right) =\frac{1}{\sqrt{2}} \left(\begin{array}{cc} 1 & -1 \\ 1 & 1 \end{array}\right) \left(\begin{array}{c} p_{\xi} \\ p_{\eta} \end{array} \right) \;, \nonumber \\
&&\left(\begin{array}{c} \sigma_x \\ \sigma_y \end{array}\right) =\frac{1}{\sqrt{2}} \left(\begin{array}{cc} 1 & -1 \\ 1 & 1 \end{array}\right) \left(\begin{array}{c} \sigma_{\xi} \\ \sigma_{\eta} \end{array}\right) 
\end{eqnarray}
and the same relations hold for the magnetic field $\bm{B}$ and the vector potential $\bm{A}$.
Then the above Hamiltonian is rewritten as
\begin{eqnarray}
&&H=\frac{1}{2m} {\bf p}^2 -p_{\xi}\sigma_{\eta}(a_D + a_R)+p_{\eta}\sigma_{\xi}(a_R - a_D) \nonumber \\
&&+\frac{1}{2} g_e \mu_B (B_{\xi}\sigma_{\xi}+ B_{\eta}\sigma_{\eta} +B_z \sigma_z) \\
&&=\frac{1}{2m} \left(p_{\xi}- \frac{\hbar}{\lambda_{\xi}}\sigma_{\eta} \right)^2 + \frac{1}{2m} \left(p_{\eta}+ \frac{\hbar}{\lambda_{\eta}}\sigma_{\xi} \right)^2 + \frac{1}{2m} p^2_z \nonumber \\
&& +\frac{1}{2} g_e \mu_B (B_{\xi}\sigma_{\xi}+ B_{\eta}\sigma_{\eta} +B_z \sigma_z)
-m(a^2_D +a^2_R)  \label{aeqq20} \\
{\rm with} \;\; && \lambda_{\xi}=\frac{\hbar}{m(a_R+a_D)} \;,\; \lambda_{\eta}=\frac{\hbar}{m(a_R-a_D)} \;,
\end{eqnarray}
where the last constant term in Eq. (\ref{aeqq20}) will be omitted hereafter. Now, in order to eliminate the spin-orbit 
coupling, we introduce the unitary transformation \cite{Aleiner}:
\begin{equation}
\tilde{H}=U^{\dag} \;H\;U \quad {\rm with} \; U=\exp[i\frac{\xi}{\lambda_{\xi}}\sigma_{\eta} - i\frac{\eta}{\lambda_{\eta}}\sigma_{\xi}] \;. \label{aeqq5}
\end{equation}
Using the formula:
\begin{equation}
\exp[i S] \;H\; \exp[-i S] =H +i [S, H]+\frac{i^2}{2!} [S, [S, H]]+ \frac{i^3}{3!} [S, [S, [S, H]]] + \cdots \;, \label{aeqq6}
\end{equation}
we calculate the terms up to the second order in $S$ and obtain
\begin{eqnarray}
&&\tilde{H}=\frac{1}{2m} {\bf p}^2 +\frac{1}{2} g_e \mu_B (B_{\xi}\sigma_{\xi}+ B_{\eta}\sigma_{\eta} +B_z \sigma_z) \nonumber \\
&&+ g_e \mu_B \left[ \left(\frac{\xi}{\lambda_{\xi}} \sigma_{\xi}+\frac{\eta}{\lambda_{\eta}} \sigma_{\eta}\right) B_z - \left(\frac{\xi}{\lambda_{\xi}} B_{\xi} +\frac{\eta}{\lambda_{\eta}} B_{\eta} \right) \sigma_z \right] \nonumber \\
&&+\frac{\hbar}{m \lambda_{\xi} \lambda_{\eta}} \left[ -L_z \sigma_z +\frac{\sigma_{\xi}}{\lambda_{\xi}}(-2\xi^2 p_{\eta}+\eta \{\xi, p_{\xi}\}) + \frac{\sigma_{\eta}}{\lambda_{\eta}}(2\eta^2 p_{\xi}- \xi \{\eta, p_{\eta}\}) \right] \;, \label{aeqq1}
\end{eqnarray}
where $\{A, B\} \equiv AB + BA$, the first line represents the single electron Hamiltonian under a magnetic field without the SO coupling, the second line the SO induced Zeeman interaction and the third line 
stands for the renormalized SO interaction. The energy level diagram of $\tilde{H}$ is totally the same as that of the original Hamiltonian $H$. Thus we can discuss the S-T$_{\pm}$ crossing behavior based on the transformed Hamiltonian $\tilde{H}$. It is important to note that owing to the unitary transformation the second and third lines of Eq. (\ref{aeqq1}) contain a smallness parameter defined by $\varepsilon \equiv \ell_t/\lambda_{\xi(\eta)}$ which is typically about $10^{-3}$, where $\ell_t$ denotes the lateral extent of the electron wavefunction. Thus we can treat these terms perturbationally.

First of all, we study the second line of Eq. (\ref{aeqq1}), which will be denoted by $H^{SO}_Z$: 
\begin{equation}
H^{SO}_Z=g_e \mu_B \left[\left(\frac{\xi}{\lambda_{\xi}} \sigma_{\xi}+\frac{\eta}{\lambda_{\eta}} \sigma_{\eta}\right) B_z - \left(\frac{\xi}{\lambda_{\xi}} B_{\xi} +\frac{\eta}{\lambda_{\eta}} B_{\eta} \right) \sigma_z \right] \;.
\end{equation}
For two electrons this part can be rewritten as
\begin{eqnarray}
&&H^{SO}_Z= g_e \mu_B \left[ \left(\frac{R_{\xi}}{\lambda_{\xi}} \Sigma_{\xi}+\frac{R_{\eta}}{\lambda_{\eta}} \Sigma_{\eta}\right) B_z - \left(\frac{R_{\xi}}{\lambda_{\xi}} B_{\xi} +\frac{R_{\eta}}{\lambda_{\eta}} B_{\eta} \right) \Sigma_z \right] \nonumber \\
&&+\frac{1}{2} g_e \mu_B \left[ \left(\frac{\xi}{\lambda_{\xi}} \sigma_{\xi}+\frac{\eta}{\lambda_{\eta}} \sigma_{\eta}\right) B_z - \left(\frac{\xi}{\lambda_{\xi}} B_{\xi} +\frac{\eta}{\lambda_{\eta}} B_{\eta} \right) \sigma_z \right] \;, \label{aeqq15}
\end{eqnarray}
where $R_{\xi}(\xi)$ and $R_{\eta}(\eta)$ denote the center-of-mass (relative) coordinates in Eq. (\ref{aeqq2}) and $\Sigma_{\xi}(\sigma_{\xi})$ and $\Sigma_{\eta}(\sigma_{\eta})$ are defined by Eq. (\ref{aeq2}). Then the matrix element between the S and T$_{\pm}$ states is given by
\begin{equation}
\langle T_{\pm}|H^{SO}_Z|S\rangle=\frac{1}{2} g_e \mu_B \langle T_{\pm}|\left(\frac{\xi}{\lambda_{\xi}} \sigma_{\xi}+\frac{\eta}{\lambda_{\eta}} \sigma_{\eta}\right) B_z - \left(\frac{\xi}{\lambda_{\xi}} B_{\xi} +\frac{\eta}{\lambda_{\eta}} B_{\eta} \right) \sigma_z|S\rangle \;.
\label{aeqq16}
\end{equation}
The direction of the magnetic field will be taken as $(\sin \theta \cos \varphi, \sin \theta \sin \varphi, \cos \theta)$ 
in the original coordinate system $(x, y, z)$ and is assumed as the direction of the spin quantization.
 Then the spin part of the matrix element is calculated as
\begin{eqnarray}
&&\langle T_{\pm}|\sigma_{\xi}|S\rangle =\mp \sqrt{2}\cos \theta \cos \varphi_- -i \sqrt{2} \sin \varphi_-  \:, \\
&&\langle T_{\pm}|\sigma_{\eta}|S\rangle =\mp \sqrt{2}\cos \theta \sin \varphi_- +i \sqrt{2} \cos \varphi_-  \:, \\
&&\langle T_{\pm}|\sigma_z|S\rangle =\pm \sqrt{2} \sin \theta \;,
\end{eqnarray}
where $\varphi_-=\varphi-\pi/4$ is the azimuth of the magnetic field in the $(\xi, \eta, z)$ 
coordinate system. The excited orbital state associated with $|T_{\pm}\rangle$ can be approximated by
\begin{equation}
e_{rel}(\xi, \eta) \propto (\xi \pm i \eta) \; g_{rel}(\xi, \eta) \;,
\end{equation}
where $g_{rel}$ is the ground state orbital of the relative coordinates.
Hereafter, we consider the case of $\xi - i \eta$ in the above equation. Then we have
\begin{equation}
\langle e_{rel}|\xi|g_{rel}\rangle=r_{10} \;,\; \langle e_{rel}|\eta|g_{rel}\rangle=i\; r_{10} \;,
\end{equation}
where $r_{10}$ is a real constant. Consequently, the matrix element is calculated as \cite{Golovach}
\begin{equation}
\langle T_{\pm}|H^{SO}_Z|S\rangle=\mp \frac{1}{\sqrt{2}} g \mu_B B r_{10} \left[ \left( \frac{1}{\lambda_{\xi}} \pm \frac{\cos \theta}{\lambda_{\eta}}\right) \cos \varphi_- +i \left( \frac{1}{\lambda_{\eta}} \pm \frac{\cos \theta}{\lambda_{\xi}}\right) \sin \varphi_- \right] \;.
\end{equation}
This result suggests that the matrix element can be made to vanish by tuning the direction of the 
magnetic field. Depending on the relative magnitude between $\lambda_{\xi}$ and $\lambda_{\eta}$, one of the factors 
\begin{equation}
\frac{1}{\lambda_{\xi}} \pm \frac{\cos \theta}{\lambda_{\eta}} \; {\rm and} \; \frac{1}{\lambda_{\eta}} \pm \frac{\cos \theta}{\lambda_{\xi}}
\end{equation}
can be made to vanish by choosing $\theta$ appropriately. Then the factor $\sin \varphi_-$ or $\cos \varphi_-$ associated with the non-zero prefactor can be made to be zero by choosing $\varphi_-$ appropriately. Usually one of the SO coupling constants $a_D$ and $a_R$ is much larger than the other and thus $|\lambda_{\xi}/\lambda_{\eta}| \cong 1$. This means that the appropriate angle $\theta$ is nearly 0 or $\pi$, indicating the $z$-directed magnetic field.

 The higher order contribution is present. For example, the second order term is given by
\begin{equation}
\sum_m \frac{\langle T_{\pm}|H^{SO}_Z|m\rangle \langle m|H^{SO}_Z|S \rangle}{E(S)- E(m)} \;,
\label{aeqq11}
\end{equation}
where $|m\rangle$ denotes appropriate intermediate states.
The typical magnitude of $H^{SO}_Z$ is estimated for $g_e \sim 1$ and $B \sim 1T$ as
\begin{equation}
H^{SO}_Z \sim g_e \mu_B B \; \frac{\ell_t}{\lambda_{SO}} \sim 5 \cdot 10^{-8} \; {\rm eV} \;,
\end{equation}
where $\lambda_{SO} \cong \lambda_{\xi} \cong \lambda_{\eta} \sim 10\mu$m and $\ell_t \sim 10$nm are assumed and then the magnitude of the above second order term is about a few times $10^{-12}$ eV because $E(S)- E(m) \sim 1$meV.

 Now we discuss the renormalized SO interaction given by the third line of Eq. (\ref{aeqq1}):
\begin{equation}  
H_{ren}=\frac{\hbar}{m \lambda_{\xi} \lambda_{\eta}} \left[ -L_z \sigma_z +\frac{\sigma_{\xi}}{\lambda_{\xi}}(-2\xi^2 p_{\eta}+\eta \{\xi, p_{\xi}\}) + \frac{\sigma_{\eta}}{\lambda_{\eta}}(2\eta^2 p_{\xi}- \xi \{\eta, p_{\eta}\}) \right] \:. \label{aeqq4}
\end{equation}
The first term $L_z \sigma_z$ in the parenthesis for two electrons can be written as
\begin{equation}
L_{1z} \sigma_{1z}+ L_{2z} \sigma_{2z}=\frac{1}{2}(L_{1z}+L_{2z}) \Sigma_z+ \frac{1}{2}(L_{1z}- L_{2z}) \sigma_z \;,
\end{equation}
where the subscript 1(2) denotes the first (second) electron.
Since the total spin operator $\Sigma_z$ conserves the magnitude of the total spin, we have
\begin{eqnarray}
&&\langle T_{\pm}|L_{1z} \sigma_{1z}+ L_{2z} \sigma_{2z}|S\rangle = \frac{1}{2} \langle T_{\pm}|(L_{1z}- L_{2z}) \sigma_z |S\rangle  \nonumber \\
&&=\langle T_{\pm}|\left[ \frac{1}{4}(\xi \Pi_{\eta} -\eta \Pi_{\xi})+R_{\xi}p_{\eta}-R_{\eta}p_{\xi} \right] \sigma_z |S\rangle \;.
\end{eqnarray}
This matrix element vanishes when the magnetic field is applied in the z direction, because the $\sigma_z$ operator does not change the magnetic quantum number. Even in the case of a tilted magnetic field, this matrix element vanishes when the orbital functions are given by
\begin{eqnarray}
&&\psi_{T_{\pm}}(R_{\xi}, R_{\eta}, \xi,\eta)=e_{cm}(R_{\xi}, R_{\eta}) g_{rel}(\xi, \eta) \; {\rm or} \; g_{cm}(R_{\xi}, R_{\eta}) e_{rel}(\xi, \eta) \;, \\
&&\psi_{S}(R_{\xi}, R_{\eta}, \xi,\eta)=g_{cm}(R_{\xi}, R_{\eta}) g_{rel}(\xi, \eta) \;,
\end{eqnarray}
where $g$ is the even parity ground state orbital and $e$ is the odd parity excited state orbital. 
Thus in the first order perturbation the $L_z \sigma_z$ term can be neglected. Now the higher order perturbation terms will be studied.
The typical magnitude of this $L_z \sigma_z$ term is estimated as
\begin{equation}
\frac{\hbar^2}{m \lambda^2_{SO}} \sim \frac{\hbar^2}{m \ell^2_t} \left(\frac{\ell_t}{\lambda_{SO}}\right)^2 \sim 10^{-8} \; {\rm eV} \;.
\end{equation}
For example, the second order perturbation term, whose expression is similar to that in Eq. (\ref{aeqq11}), contributes an amount of the order of $10^{-13}$ eV. In the higher order perturbation series the magnitude becomes even smaller and the contribution of the $L_z \sigma_z$ term to the S-T$_{\pm}$ anticrossing gap would be of the order of $10^{-13}$ eV.

Now we discuss the residual terms of the renormalized SO interaction in Eq. (\ref{aeqq4}). These terms for two electrons can be rewritten in terms of the center-of-mass coordinates and the relative coordinates and again only the part associated with the relative coordinates contributes to the matrix element between the singlet state $|S\rangle$ and the triplet states $|T_{\pm}\rangle$. The relevant part is given by
\begin{eqnarray}
&&H'_{ren}=\frac{\hbar}{m \lambda_{\xi}\lambda_{\eta}} \times \nonumber \\
&&\Bigl[ \frac{\sigma_{\xi}}{\lambda_{\xi}} \left(2(R_{\xi}R_{\eta}+\frac{1}{4}\xi\eta) p_{\xi} +\frac{1}{2}(R_{\eta}\xi+ R_{\xi}\eta) \Pi_{\xi} 
-2(R^2_{\xi} +\frac{1}{4}\xi^2) p_{\eta}-R_{\xi}\xi \Pi_{\eta} -\frac{i\hbar}{2} \eta \right) \nonumber \\
&& -\frac{\sigma_{\eta}}{\lambda_{\eta}} \left( 2(R_{\xi} R_{\eta}+\frac{1}{4}\xi\eta) p_{\eta} +\frac{1}{2} (R_{\eta} \xi+R_{\xi}\eta) \Pi_{\eta}
-2(R^2_{\eta} +\frac{1}{4}\eta^2) p_{\xi}-R_{\eta}\eta \Pi_{\xi} -\frac{i\hbar}{2} \xi \right) \Bigr] \;. \nonumber \\
&& \quad
\end{eqnarray}
The typical magnitude of this term is estimated as
\begin{equation}
\frac{\hbar^2 \ell_t}{m \lambda^3_{SO}} \sim \frac{\hbar^2}{m \ell^2_t} \left(\frac{\ell_t}{\lambda_{SO}}\right)^3 \sim 10^{-11} \; {\rm eV} \;.
\end{equation}
In the higher order perturbation series, the contribution becomes much smaller. Thus the contribution from the renormalized SO interaction to the S-T$_{\pm}$ anticrossing gap is of the order of $10^{-11}$ eV. 

Summarizing the above arguments within the linear-in-momentum SO coupling, we can conclude that the S-T$_{\pm}$ anticrossing gap is mainly contributed by the SO induced Zeeman interaction $H^{SO}_Z$ and this contribution can be eliminated in the first order by tuning the direction of the magnetic field, namely, by the magic angle tuning. However, the actual anticrossing gap is determined by other terms and the higher order perturbation terms of $H^{SO}_Z$ and is of the order of $10^{-12} \sim 10^{-11}$ eV. This magnitude is comparable to the hf interaction energy in the case of two delocalized electrons in a QD as discussed in Sec. IV.

In general, the cubic-in-momentum spin-orbit term is present and more detailed arguments are necessary. The Dresselhaus SO term is originally given as
\begin{eqnarray}
&&V_D= \gamma \; \bm{\Lambda} \cdot \bm{\sigma} \\
{\rm with} \; && \Lambda_x=k_x (k^2_y - k^2_z) \;, \; \Lambda_y=k_y (k^2_z - k^2_x) \;, \; \Lambda_z=k_z (k^2_x - k^2_y) \;, 
\end{eqnarray}
where $\hbar k_i=p_i \; (i=x, y, z)$ including the vector potential due to a magnetic field. In the two-dimensional limit, we usually take the matrix element
\begin{equation}
V_D^{(1)}= \langle \phi(z)|V_D|\phi(z)\rangle= \gamma \langle \phi(z)| k^2_z |\phi(z)\rangle (- k_x \sigma_x + k_y \sigma_y) 
\end{equation}
with the ground state orbital $\phi(z)$ in the $z$-direction and put as
\begin{equation}
=a_D (p_x \sigma_x - p_y \sigma_y) \quad
{\rm with} \quad a_D=- \frac{\gamma}{\hbar} \langle \phi(z)| k^2_z |\phi(z)\rangle  \label{aeq4}
\end{equation}
and call this the linear-in-momentum SO term which is already included in Eq. (\ref{so}). Thus, in general, we can put
\begin{eqnarray}
&&V_D=V_D^{(1)}+V_D^{(3)} \\
{\rm with} \; && V_D^{(3)}= \gamma (k_x k^2_y \sigma_x -k_y k^2_x \sigma_y) \;,
\end{eqnarray}
where $V_D^{(3)}$ is called the cubic-in-momentum Dresselhaus SO coupling term. In the $(\xi, \eta, z)$ coordinate system this term is rewritten as
\begin{equation}
V_D^{(3)}= \frac{\gamma}{4\hbar^3} \left[ (\{p^2_{\xi}, p_{\eta}\}-2 p^3_{\eta})\sigma_{\xi} + (\{p^2_{\eta}, p_{\xi}\}-2 p^3_{\xi})\sigma_{\eta} \right] \;.
\end{equation}
After the unitary transformation in Eq. (\ref{aeqq5}) we have
\begin{eqnarray}
&&U^{\dag}\;V_D^{(3)} \;U = V_D^{(3)} + \frac{\gamma}{2\hbar^2 \lambda_{\xi}} (p^2_{\eta} -3 p^2_{\xi})
-\frac{\gamma}{2\hbar^2 \lambda_{\eta}} (p^2_{\xi} -3 p^2_{\eta}) \nonumber \\
&&+ \frac{\gamma}{4\hbar^3} \Bigl[ \frac{2i}{\lambda_{\xi}}(-\hbar\{p_{\xi}, p_{\eta}\} +i(\{p^2_{\xi}, p_{\eta}\} -2 p^3_{\eta}) \xi )
 + \frac{2i}{\lambda_{\eta}}(-\hbar\{p_{\xi}, p_{\eta}\} +i(\{p^2_{\eta}, p_{\xi}\} -2 p^3_{\xi}) \eta ) \Bigr] \sigma_z + \cdots  \nonumber \\
&&=V_D^{(3)} + V_D^{(3)ren} + \cdots \;, \label{aeqq18}
\end{eqnarray}
where the results are given up to the first order in the expansion of Eq. (\ref{aeqq6}) and $V_D^{(3)ren}$
is defined by the terms on the right hand side other than $V_D^{(3)}$.  
Unfortunately, $V_D^{(3)}$ cannot be eliminated. But the renormalized terms $V_D^{(3)ren}$ are smaller in 
magnitude than the original terms due to the smallness parameter $\varepsilon \equiv \ell_t/\lambda_{SO} \sim 10^{-3}$. Furthermore, those terms do not contribute to the matrix element between the singlet state $|S\rangle$ and the triplet states $|T_{\pm}\rangle$ because they do not change the magnetic quantum number.
Thus we have to consider the effect of the original $V_D^{(3)}$ on the S-T$_{\pm}$ anticrossing gap.

The coupling $V_D^{(3)}$ for two electrons in the original coordinate system can be rewritten like in Eq. (\ref{aeq1}) as
\begin{eqnarray}
&&V_D^{(3)}=\frac{\gamma}{\hbar^3} \Bigl[ \{ \frac{1}{2} \Pi_x(p^2_y+\frac{1}{4}\Pi^2_y)+p_x p_y \Pi_y \} \Sigma_x
+\{ p_x(p^2_y+\frac{1}{4}\Pi^2_y)+ \frac{1}{2} \Pi_x \Pi_y p_y \} \sigma_x \Bigr] \nonumber \\
&&- \frac{\gamma}{\hbar^3} \Bigl[ \{ \frac{1}{2} \Pi_y(p^2_x+\frac{1}{4}\Pi^2_x)+p_x p_y \Pi_x \} \Sigma_y
- \{p_y(p^2_x+\frac{1}{4}\Pi^2_x)+ \frac{1}{2} \Pi_x \Pi_y p_x \} \sigma_y \Bigr] \;. \label{aeqq8}
\end{eqnarray}
The operators $\Sigma_x$ and $\Sigma_y$ do not change the magnitude of the total spin and thus they do not contribute to the matrix element $\langle S|V_D^{(3)}|T_{\pm} \rangle$. Accordingly, we find
\begin{eqnarray}
&&\langle S|V_D^{(3)}|T_{\pm} \rangle=\mp \sqrt{2} \frac{\gamma}{\hbar^3} \langle g_{cm} g_{rel}|p_x(p^2_y+\frac{1}{4}\Pi^2_y)+ \frac{1}{2} \Pi_x \Pi_y p_y |g_{cm} e_{rel}\rangle \nonumber \\
&& +i \sqrt{2} \frac{\gamma}{\hbar^3} \langle g_{cm} g_{rel}|p_y(p^2_x+\frac{1}{4}\Pi^2_x)+ \frac{1}{2} \Pi_x \Pi_y p_x |g_{cm} e_{rel}\rangle \;.
\end{eqnarray}
The linear terms with respect to the $p_x$ and $p_y$ can be included in Eq. (A1) to renormalize the linear-in-momentum SO interaction and the unitary transformation in Eq. (A26) can be redefined. Consequently, we have
\begin{eqnarray}
&&\langle S|V_D^{(3)}|T_{\pm} \rangle=\mp \sqrt{2} \frac{\gamma}{\hbar^3} \langle g_{rel}|p_x p^2_y \mp i \; p_y p^2_x| e_{rel}\rangle \;.
\end{eqnarray}
In order to estimate these matrix elements, we have to symmetrize the operators as
\begin{equation}
p_x p^2_y \longrightarrow \frac{1}{2} \{p_x, p^2_y \} =\frac{1}{2} (p_x p^2_y + p^2_y p_x) \;, \; p_y p^2_x \longrightarrow \frac{1}{2} \{p_y, p^2_x \}=\frac{1}{2} (p_y p^2_x + p^2_x p_y) \;,
\end{equation}
because $p_x$ and $p_y$ do not commute. The odd-parity excited orbital state of the relative motion can be approximated as
\begin{equation}
e^{\pm}_{rel}(x,y)=\frac{1}{\ell_t} (x \pm iy) \; g_{rel}(x,y)\; {\rm with} \; \;g_{rel}(x,y)=\frac{1}{\ell_t \sqrt{\pi}} \exp[-\frac{1}{2\ell^2_t}(x^2+y^2)] \;,
\end{equation}
where the $z$-coordinate part is omitted. Then we have
\begin{eqnarray}
&&\langle g_{rel}|\frac{1}{2} (p_x p^2_y + p^2_y p_x)| e^{\pm}_{rel}\rangle =\frac{i \hbar^3}{2 \ell_t} \Bigl[-\frac{1}{2 \ell^2_t} -\frac{\ell^2_t}{32 \ell^4_B} \mp \frac{1}{8 \ell^2_B} \mp \frac{\ell^4_t}{128 \ell^6_B} \Bigr] \;, \\
&& \langle g_{rel}|\frac{1}{2} (p_y p^2_x + p^2_x p_y)| e^{\pm}_{rel}\rangle =\frac{ \hbar^3}{2 \ell_t} \Bigl[\pm \frac{1}{2 \ell^2_t} \pm \frac{\ell^2_t}{32 \ell^4_B} + \frac{1}{8 \ell^2_B} + \frac{\ell^4_t}{128 \ell^6_B} \Bigr]
\\
{\rm with} && \ell_B=\sqrt{\frac{\hbar c}{e B}} \;,
\end{eqnarray}
where $\ell_B$ is the magnetic length. The matrix element is calculated as
\begin{eqnarray}
&&\langle S|V_D^{(3)}|T_{+} \rangle=- \sqrt{2} \frac{\gamma}{\hbar^3} \langle g_{rel}|\frac{1}{2} \Bigl(\{p_x, p^2_y\} - i \; \{p_y, p^2_x\} \Bigr)| e_{rel}\rangle  \nonumber \\
&&=\left\{ \begin{array}{c} \frac{i\sqrt{2}\gamma}{\ell_t} \Bigl(\frac{1}{2 \ell^2_t} + \frac{\ell^2_t}{32 \ell^4_B} \Bigr) \Bigl(1+\frac{\ell^2_t}{4\ell^2_B} \Bigr) \quad {\rm for} \quad e_{rel}(x,y)=e^+_{rel}(x,y) \\
\hspace{3cm} 0 \hspace{2.3cm} {\rm for} \quad e_{rel}(x,y)=e^-_{rel}(x,y) \;.
\end{array}
\right.
\end{eqnarray}
In the same way, we find that $\langle S|V_D^{(3)}|T_{-} \rangle=0$ for $e_{rel}(x,y)=e^+_{rel}(x,y)$. In general, the S-T$_+$ and S-T$_-$ crossing occur at different magnitudes of the magnetic field and we have to select an appropriate crossing point to realize the efficient electron-nuclear spin QST.
In any case, by choosing the right S-T crossing point, we can eliminate the first order term of the cubic-in-momentum Dresselhaus SO coupling. 

Now we consider the higher order perturbation terms with respect to $V_D^{(3)}$. The next third-order terms are given by
\begin{eqnarray}
(a) &&\sum_{S', T'_{\pm}} \frac{ \langle S|V_D^{(3)}|T'_{\pm} \rangle \langle T'_{\pm}|V_D^{(3)}|S' \rangle \langle S'|V_D^{(3)}|T_{\pm} \rangle}{(E(T_{\pm})-E(S'))(E(T_{\pm})-E(T'_{\pm})} \;,  \label{aeqq12} \\
{\rm and}\quad  (b) && \sum_{T_0, T'_{\pm}} \frac{ \langle S|V_D^{(3)}|T'_{\pm} \rangle \langle T'_{\pm}|V_D^{(3)}|T_0 \rangle \langle T_0|V_D^{(3)}|T_{\pm} \rangle}{(E(T_{\pm})-E(T_0))(E(T_{\pm})-E(T'_{\pm}))} \;, \label{aeqq13}
\end{eqnarray}
where $\dn{S'}$ and $\dn{T'_{\pm}}$ are the singlet and triplet states different from $\dn{S}$ and $\dn{T_{\pm}}$. In the matrix elements of $\langle T'_{\pm}|V_D^{(3)}|T_0 \rangle$ and $\langle T_0|V_D^{(3)}|T_{\pm} \rangle$, the terms proportional to $\Sigma_x$ and $\Sigma_y$ in Eq. (\ref{aeqq8}) contribute. The general eigenstates can be written as
\begin{eqnarray}
&&|T_+(T_-)\rangle=\psi_{cm}({\bf R})\phi_{rel}({\bf r}) \; \alpha(\xi_1)\alpha(\xi_2) \;(\beta(\xi_1)\beta(\xi_2)) \;,\\
&&|T_0\rangle=\psi'_{cm}({\bf R})\phi_{rel}({\bf r}) \frac{1}{\sqrt{2}}(\alpha(\xi_1)\beta(\xi_2)+ \beta(\xi_1)\alpha(\xi_2)) \;,
\end{eqnarray}
where $\phi_{rel}$ is an odd-parity excited state of the relative motion and $\psi_{cm}$ and $\psi'_{cm}$ are appropriate orbital functions of the center-of-mass motion. 
 The magnitude of the matrix elements in Eqs. (\ref{aeqq12}) and (\ref{aeqq13}) is typically of the order of
\begin{equation}
\langle S|V_D^{(3)}|T_{+} \rangle \simeq \frac{\gamma}{\ell_t^3} \sim 1 \;\mu{\rm eV}
\end{equation}
for $\gamma \sim 10$ eV(\AA$^3$) \cite{Winkler} and $\ell_t \sim$ 20 nm. Then the magnitude of terms in Eqs. (\ref{aeqq12}) and (\ref{aeqq13}) is estimated as
\begin{equation}
 \frac{1 \;\mu{\rm eV} \times 1 \;\mu{\rm eV} \times 1 \;\mu{\rm eV}}{\Delta \;\Delta'} \sim  10^{-12} {\rm eV} \;, \label{aeqq10}
\end{equation}
where $\Delta$ and $\Delta'$ are the orbital energy difference of the order of 1 meV.
The higher order perturbation terms with respect to $V^{(3)}_D$ become smaller furthermore.
Finally, we discuss the contribution from $V^{(3)ren}_{D}$ in Eq. (\ref{aeqq18}) which is smaller than $V^{(3)}_D$ by the smallness parameter $\varepsilon$. The first order term $\langle S|V^{(3)ren}_{D}|T_{\pm}\rangle$ 
vanishes because $V^{(3)ren}_{D}$ does not change the magnetic quantum number. In the higher order perturbation terms, $V^{(3)ren}_{D}$ contributes in combination with other interaction Hamiltonians, e.g.,
$V^{(3)}_{D}$. For example, the second order perturbation terms have the magnitude of about $10^{-12}$ eV. The magnitude of higher order terms becomes smaller furthermore.

Consequently, we can summarize that the S-T$_{\pm}$ anticrossing gap due to the Dresselhaus and Rashba SO interactions would be of the order of $10^{-12} \sim 10^{-11}$ eV and is of the same order of magnitude as the hf coupling energy for the case of two delocalized electrons in a QD. This means that our protocol for the muclear spin quantum memory is not effective in the 
case of two delocalized electrons in a QD.

So far we have considered the case of two delocalized electrons in a QD with an isotope atom of the host material. In the respect of the electron-nuclear spin QST time, the single electron charged QD with a donor impurity having the nuclear spin is more favorable. Especially favorable is the case where one electron is delocalized within the QD and the other electron is strongly bound to the ionized donor, as discussed in Sec. IV. In this case the hf coupling energy is of the order of $10^{-7} \sim 10^{-6}$ eV. Now we examine the S-T$_{\pm}$ anticrossing gap due to the SO interactions. The most relevant term after the unitary transformation in Eq. (\ref{aeqq1}) is the SO induced Zeeman interaction $H^{SO}_Z$:
\begin{equation}
H^{SO}_Z=g_e \mu_B \left[ \left(\frac{\xi}{\lambda_{\xi}} \sigma_{\xi}+\frac{\eta}{\lambda_{\eta}} \sigma_{\eta}\right) B_z - \left(\frac{\xi}{\lambda_{\xi}} B_{\xi} +\frac{\eta}{\lambda_{\eta}} B_{\eta} \right) \sigma_z \right] \:. 
\end{equation}
This interaction for two electrons can be rewritten in terms of the total spin operators and the difference spin operators as given in Eq. (\ref{aeqq15})
and the matrix element between the singlet state $|S\rangle$ and the triplet state $|T_{\pm}\rangle$ is given by Eq. (\ref{aeqq16}).
The orbital parts of the singlet and triplet states can be approximated by
\begin{eqnarray}
&&|\Psi_S\rangle=\frac{1}{\sqrt{2}} (\phi_a({\bf r_1}) \phi_b({\bf r_2})+\phi_b({\bf r_1}) \phi_a({\bf r_2})) \;,\\
&&|\Psi_T \rangle=\frac{1}{\sqrt{2}}[\phi_a({\bf r_1}) \phi_b({\bf r_2})- \phi_b({\bf r_1}) 
\phi_a({\bf r_2})]  \;,
\end{eqnarray}
where $\phi_a(\rb)$ is the delocalized orbital in the QD and $\phi_b(\rb)$ is the strongly localized donor-bound orbital.
Hereafter the overlap integral between $\phi_a(\rb)$ and $\phi_b(\rb)$ will be neglected because the donor is doped in the peripheral region of the QD. Then we have
\begin{eqnarray}
&&\langle \Psi_T|\xi|\Psi_S\rangle=\langle \Psi_T|\xi_1-\xi_2|\Psi_S\rangle \cong \langle \phi_a|\xi|\phi_a\rangle- \langle \phi_b|\xi|\phi_b\rangle \;, \\
&&\langle \Psi_T|\eta|\Psi_S\rangle=\langle \Psi_T|\eta_1-\eta_2|\Psi_S\rangle \cong \langle \phi_a|\eta|\phi_a\rangle- \langle \phi_b|\eta|\phi_b\rangle \;.
\end{eqnarray}
The center of the laterally symmetric confinement potential is chosen at the origin and the position of the donor atom is taken as $(d, d, 0)$ in the $(\xi, \eta, z)$ coordinate system, leading to
\begin{equation}
\langle \Psi_T|\xi_1-\xi_2|\Psi_S\rangle \cong -d \;, \; \langle \Psi_T|\eta_1-\eta_2|\Psi_S\rangle \cong -d \;.
\end{equation}
Then we obtain
\begin{equation}
\langle T_{\pm}|H^{SO}_Z|S\rangle =\pm \frac{g_e \mu_B B d}{\sqrt{2}} \left[\frac{\cos \varphi_-}{\lambda_{\xi}}+\frac{\sin \varphi_-}{\lambda_{\eta}} \pm i\cos \theta \left(\frac{\sin \varphi_-}{\lambda_{\xi}}-\frac{\cos \varphi_-}{\lambda_{\eta}} \right) \right] \;.
\end{equation}
This result suggests that the matrix element vanishes when $\theta=\pi/2$ and $\tan \varphi_-=-\lambda_{\eta}/\lambda_{\xi}$. However, this means the in-plane magnetic field and does not conform to the previous arguments developed for the case of a longitudinal (z-directed) magnetic field which is a prerequisite for the S-T level crossing. Thus the magic angle tuning is not possible here. Instead, in order to reduce the matrix element, we have to use the group IV elemental semiconductors in which the Dresselhaus SO terms are absent and the magnitude of the Rashba SO term might be reduced very much by careful tuning of the strain fields and the electric field. If we can achieve $\ell_t/\lambda_{\xi}, \;\ell_t/\lambda_{\eta} \sim 10^{-4}$, the above matrix element would be about $10^{-8}$ eV and be smaller than the hf coupling energy. At the same time, the contribution from the higher order perturbation series is much smaller.

The arguments on the contribution from the renormalized SO interaction $H_{ren}$ in Eq. (\ref{aeqq4}) can be developed in the same way as in the case of two delocalized electrons in a QD and the relevant matrix element is of the order of $10^{-12} \sim 10^{-11}$ eV.

Finally, we examine the cubic-in-momentum Dresselhaus SO interaction $V^{(3)}_D$, although in the group IV elemental semiconductors this interaction is absent.
As discussed before, $V^{(3)}_D$ cannot be eliminated by the unitary transformation and should be considered in the original form. $V^{(3)}_D$ for two electrons is written as
\begin{eqnarray}
&&V_D^{(3)}=\frac{\gamma}{4 \hbar^3} \Bigl[ (\{p_{1x}, p^2_{1y}\}+ \{p_{2x}, p^2_{2y}\}) \Sigma_x - (\{p_{1y}, p^2_{1x}\}+ \{p_{2y}, p^2_{2x}\}) \Sigma_y \nonumber \\
&&+ (\{p_{1x}, p^2_{1y}\}- \{p_{2x}, p^2_{2y}\}) \sigma_x - (\{p_{1y}, p^2_{1x}\}- \{p_{2y}, p^2_{2x}\}) \sigma_y \Bigr] \;,
\end{eqnarray}
where the momentum operators are symmetrized. Using the functions in Eqs. (\ref{phia}) and (\ref{phib}), we can show that
\begin{equation}
\langle S|V^{(3)}_D|T_{\pm} \rangle =0
\end{equation}
by elementary calculations. The higher order perturbation terms with respect to $V^{(3)}_D$ are present but the same arguments as in Eq. (\ref{aeqq10}) hold and their contribution is of the order of $10^{-12}$ eV. Finally, we discuss the contribution from $V^{(3)ren}_{D}$ in Eq. (\ref{aeqq18}) which is smaller than $V^{(3)}_D$ by the smallness parameter $\varepsilon$. The same arguments below Eq. (\ref{aeqq10}) can be applied and the typical magnitude is estimated to be of the order of $10^{-12}$ eV.

Consequently, in the case of the single electron charged QD doped with a donor impurity having the nuclear spin, the S-T$_{\pm}$ anticrossing gap due to the SO interactions can be neglected compared with the hf interaction energy. To achieve this situation, the group IV elemental semiconductors with the reduced Rashba SO interaction is favorable, which might be realized by careful tuning of the strain fields and the electric field. In these systems our protocol for the nuclear spin quantum memory would be fully effective.

\section{Integrals appearing in the expression of $J$}

The relevant integrals appearing in the expression of $J$ in Eq. (\ref{j}) are given here.
\begin{align}
\av{\phi_a|V_b|\phi_a}=&-R \frac{1}{\sqrt{\pi}}\frac{1}{\ell_t^2\ell_z}\int_{-\infty}^{\infty}
ds~(s^2+\frac{1}{\ell_t^2})^{-1}(s^2+\frac{1}{\ell_z^2})^{-1/2}\exp[\frac{d^2}{\ell_t^4 (s^2+1/\ell_t^2)}-\frac{d^2}{\ell_t^2}] \;, \\
\av{\phi_b|V_a|\phi_b}=&\frac{(\hbar\omega_0)^2}{2 R} (\frac{1}{\alpha}+d^2)+ \frac{(\hbar\omega_z)^2}{2 R} \frac{1}{2\beta} \;,  \\
\av{\phi_a|V_b|\phi_b}=&-R \frac{1}{\sqrt{\pi}}\sqrt{\frac{\alpha \beta^{1/2}}{\ell_t^2 \ell_z}}\nonumber \\ & \times\int_{-\infty}^{\infty}ds~(s^2+\frac{\alpha}{2}+\frac{1}{2\ell_t^2})^{-1}(s^2+\frac{\beta}{2}+\frac{1}{2\ell_z^2})^{-1/2}
\exp[\frac{-q^2+d^2/\ell_t^4}{4(s^2+\frac{\alpha}{2}+\frac{1}{2\ell_t^2})}-\frac{d^2}{2\ell_t^2}] \;, \\
V_X=&\av{\phi_a(\rb_1)\phi_b(\rb_2)|\frac{e^2}{\kappa|\rb_1-\rb_2|}|\phi_a(\rb_2)\phi_b(\rb_1)}=R \frac{\alpha \beta^{1/2}}{\ell_t^2\ell_z}\frac{1}{\sqrt{\pi}}\nonumber \\ 
&\times (\beta+\frac{1}{\ell_z^2})^{-1/2}(\alpha+\frac{1}{\ell_t^2})^{-1}\exp[\frac{d^2}{\ell_t^4 (\alpha+1/\ell_t^2)}-\frac{d^2}{\ell_t^2}] \nonumber\\ 
&\times \int_{-\infty}^{\infty}ds (s^2+\frac{1}{4\ell_t^2}+\frac{\alpha}{4})^{-1}(s^2+\frac{1}{4\ell_z^2}+\frac{\beta}{4})^{-1/2} \exp[-\frac{q^2}{4(s^2+\frac{1}{4\ell_t^2}+\frac{\alpha}{4})}] \;, \\
V_d=&\av{\phi_a(\rb_1)\phi_b(\rb_2)|\frac{e^2}{\kappa|\rb_1-\rb_2|}|\phi_a(\rb_1)\phi_b(\rb_2)}\simeq \av{\phi_a(\rb_1)|\frac{e^2}{\kappa |\rb_1-d \hat{x}|}|\phi_a(\rb_1)}\label{vd}\nonumber\\
=& R \frac{1}{\ell_t^2\ell_z}\frac{1}{\sqrt{\pi}}\int_{-\infty}^{\infty}ds\; (s^2+1/\ell_z^2)^{-1/2}(s^2+1/\ell_t^2)^{-1}\exp[\frac{d^2}{\ell_t^4 (s^2+1/\ell_t^2)}] \exp[-d^2/\ell_t^2] \;, \\
S=&\sqrt{\frac{\alpha \beta^{1/2}}{\ell_t^2\ell_z}}(\frac{\alpha}{2}+\frac{1}{2\ell_t^2})^{-1}(\frac{\beta}{2}+\frac{1}{2 \ell_z^2})^{-1/2} \exp[-\frac{d^2}{2\ell_t^2}+\frac{d^2/\ell_t^4-q^2}{2(\alpha+1/(\ell_t^2))}] \;, \\
q=&\frac{eBd}{2\hbar c} a^2_B \;,
\end{align}
where $R, a_B, \ell_t, \ell_z, d, \alpha$ and $\beta$ are defined in Sec. III and the length variables $\ell_t, \ell_z$ and $d$ are scaled by $a_B$ to be dimensionless and the parameters $q$ and $s$ are also dimensionless.
In the calculation, the transformation is employed for the Coulomb potential:
\[\frac{1}{r}=\frac{1}{\sqrt{\pi}}\int_{-\infty}^{\infty}ds~\exp[-s^2r^2].
\]
 In the above, only the direct Coulomb term $V_d$ is approximated by assuming that the QD confinement is much weaker than the donor atom confinement, i.e., $\ell_t,\ell_z \gg a_B/\sqrt{\alpha}, a_B/\sqrt{\beta}$.

\section{Quantum state transfer between an electron and a nuclear spin in the general mixed state}

 Here we examine the quantum state transfer between an electron and a nuclear spin in the general mixed state, assuming
\begin{equation}
\rho_n=\gamma |\uparrow\rangle_n \;_n\langle\uparrow| + \delta |\downarrow\rangle_n \;_n\langle\downarrow| +\epsilon |\uparrow\rangle_n \;_n\langle\downarrow| + \epsilon^* |\downarrow\rangle_n \;_n\langle\uparrow|\;, \label{aeq12}
\end{equation}
where $\gamma$ and $\delta$ are arbitrary non-negative real constants satisfying $\gamma+\delta=1$, $\epsilon$ is an arbitrary complex constant and the suffix $n$ attached to the bra and ket vectors indicates the nucleus. The electron spin is prepared in a pure state given by
\begin{equation}
|\psi\rangle_e=\alpha |\uparrow\rangle_e +\beta |\downarrow\rangle_e \;,
\end{equation} 
where $\alpha$ and $\beta$ are arbitrary complex constants normalized as $|\alpha|^2+|\beta|^2=1$ and the suffix $e$ indicates the electron. Then the time evolution of the density matrix of the electron-nucleus coupled system is described by
\begin{eqnarray}
&&\rho(t)= e^{-\frac{i}{\hbar}Ht} \; \rho(0) \; e^{\frac{i}{\hbar}Ht} \\
{\rm with} &&\rho(0)=|\psi\rangle_e \;_e\langle\psi| \otimes \rho_n 
=\left( \begin{array}{cccc}
|\alpha|^2 \gamma & |\alpha|^2 \epsilon & \alpha \beta^* \gamma & \alpha \beta^* \epsilon \\
|\alpha|^2 \epsilon^* & |\alpha|^2 \delta & \alpha \beta^* \epsilon^* & \alpha \beta^* \delta \\
\alpha^* \beta \gamma & \alpha^* \beta \epsilon & |\beta|^2 \gamma & |\beta|^2 \epsilon \\
\alpha^* \beta \epsilon^* & \alpha^* \beta \delta & |\beta|^2 \epsilon^* & |\beta|^2 \delta 
\end{array} \right) \;,
\end{eqnarray}
where the bases of the matrix representation are arranged in the order of $|\uparrow\rangle_e |\uparrow\rangle_n \;,$ $|\uparrow\rangle_e |\downarrow\rangle_n \;,$ $|\downarrow\rangle_e |\uparrow\rangle_n$ and $|\downarrow\rangle_e |\downarrow\rangle_n$. Then, using the expression of the hf interaction in Eq. (\ref{ish}), we have
\begin{equation}
e^{-\frac{i}{\hbar}Ht}=\left( \begin{array}{cccc}
e^{-i At/(4\hbar)} & 0 & 0 & 0 \\
0 & e^{i At/(4\hbar)} \;\cos \frac{At}{2\hbar} & -i \;e^{i At/(4\hbar)} \;\sin \frac{At}{2\hbar} & 0 \\
0 & -i \;e^{i At/(4\hbar)} \;\sin \frac{At}{2\hbar} & e^{i At/(4\hbar)} \;\cos \frac{At}{2\hbar} & 0 \\
0 & 0 & 0 & e^{-i At/(4\hbar)} 
\end{array} \right) 
\end{equation}
and
\begin{eqnarray}
&&\rho(t=\pi \hbar/A)=\left( \begin{array}{cccc}
|\alpha|^2 \gamma & \alpha \beta^* \gamma & |\alpha|^2 \epsilon & \alpha \beta^* \epsilon \\
\alpha^* \beta \gamma & |\beta|^2 \gamma & \alpha^* \beta \epsilon & |\beta|^2 \epsilon \\
|\alpha|^2 \epsilon^* & \alpha \beta^* \epsilon^* & |\alpha|^2 \delta & \alpha \beta^* \delta \\
\alpha^* \beta \epsilon^* & |\beta|^2 \epsilon^* & \alpha^* \beta \delta & |\beta|^2 \delta
\end{array} \right) \nonumber \\
&&=(\gamma |\uparrow\rangle_e \;_e\langle \uparrow| + \delta |\downarrow\rangle_e \;_e\langle \downarrow|
+\epsilon |\uparrow\rangle_e \;_e\langle\downarrow| + \epsilon^* |\downarrow\rangle_e \;_e\langle\uparrow|) \otimes |\psi\rangle_n \;_n\langle \psi| \nonumber \\
&&= \rho_e \otimes |\psi\rangle_n \;_n\langle \psi| \\ 
{\rm with} && |\psi\rangle_n=\alpha |\uparrow\rangle_n +\beta |\downarrow\rangle_n \;,
\end{eqnarray}
where $\rho_e$ is the same density matrix for the electron as in Eq. (\ref{aeq12}). 
This means that the QST or exchange of states between the electron and the nuclear spin is accomplished for the general mixed state of the nuclear spin.

\section{Hyperfine coupling constant in a localized electron system of ZnSe:F}

The hf coupling between a donor electron and a donor nucleus is given by the Fermi contact interaction \cite{Abragam96} as discussed in Sec. IV: 
\begin{equation}
 V_{hf}=\frac{8\pi}{3} g_e \mu_B g_n \mu_n \; {\bf S\cdot I} \; \delta({\bf r}-{\bf R}) \;,
\end{equation}
where ${\bf S (I)}$ is the dimensionless spin angular momentum operator for the donor electron (nucleus), $\mu_B (\mu_n)$ the Bohr (nuclear) magneton, $g_e (g_n)$ the g-factor of the donor electron (nucleus), and ${\bf r}({\bf R})$ denotes the position vector of the donor electron (nucleus). The donor-bound electron state can be represented by
\begin{eqnarray}
&&\Psi({\bf r})=F({\bf r}) u({\bf r})  \\
{\rm with} &&\frac{1}{v_0}\int_{v_0} d{\bf r} |u({\bf r})|^2 =1 \;, \; \int d{\bf r} |F({\bf r})|^2 =1 \;, \label{apn}
\end{eqnarray}
where $u({\bf r})$ is the Bloch function of the relevant conduction band normalized in the volume $v_0$ of a unit cell and $F({\bf r})$ is the envelope function. Then the hf coupling Hamiltonian is given by
\begin{equation}
\langle\Psi|V_{hf}|\Psi\rangle= A \; {\bf S\cdot I} \quad {\rm with} \quad A=\frac{8\pi}{3} g_e \mu_B g_n \mu_n \; |F({\bf R}) u({\bf R})|^2 \;. \label{aphf}
\end{equation}
Now we infer the coupling constant $A$ for the ZnSe:F in comparison with the case of Si:P based on Eq. (\ref{aphf}). The squared amplitude of the envelope function can be roughly estimated by
\begin{equation}
|F({\bf R})|^2 \simeq \frac{1}{a_B^3} \label{apf}
\end{equation}
in terms of the donor electron Bohr radius given by
\begin{equation}
a_B=\frac{\hbar^2 \kappa}{m^* e^2} \;,
\end{equation}
where $m^*$ is the isotropic effective mass of the donor electron and $\kappa$ is the dielectric constant of the host material. In the case of the anisotropic effective mass like in Si, Eq. (\ref{apf}) is modified as
\begin{equation}
|F({\bf R})|^2 \simeq \frac{1}{a_t^2 a_l} \;, 
\end{equation}
where $a_{t (l)}$ is the donor electron Bohr radius in the transverse (longitudinal) direction. 
Furthermore, it will be simply assumed that the Bloch function amplitude $|u({\bf R})|$ is not much different between Si and ZnSe.
Then we can compare the coupling constant $A$ between ZnSe:F and Si:P:
\begin{equation}
\frac{A({\rm ZnSe:F})}{A({\rm Si:P})}=\frac{g_e({\rm ZnSe}) g_n({\rm F})}{g_e({\rm Si}) g_n({\rm P})} \; \frac{a_t^2({\rm Si})\; a_l({\rm Si})}{a_B^3({\rm ZnSe})} \;.
\end{equation}
 For Si, the g-factor of the donor electron is $g_e({\rm Si}) \simeq 2.0$, the nuclear g-factor of the P atom is $g_n({\rm P})=1.13$, and the donor electron Bohr radius is $\simeq 3.37 (0.695)$ nm in the transverse (longitudinal) direction. On the other hand, for ZnSe, the nuclear g-factor of the F atom is $g_n({\rm F})=2.63$, and the isotropic donor electron Bohr radius $a_B$ is $\simeq 3.01$nm. The g-factor of the donor electron in ZnSe:F is not well known but it may be reasonable to guess that $g_e({\rm ZnSe}) \simeq 2.0$, because the measured g-factors of the donor-electron of many other impurities in ZnSe are accumulating around 2.0. Then the ratio of $A$ is about 0.672
and we infer $A({\rm ZnSe:F}) \simeq 78.6$ MHz from the value $A({\rm Si:P}) \simeq 117$ MHz.

\newpage


\end{document}